\documentclass[mathpazo]{journal}

\usepackage[linesnumbered,ruled,vlined]{algorithm2e}
\usepackage{float}
\usepackage{multirow}
\newcommand{\R}{\mathbb{R}}
\newcommand{\sita}{\boldsymbol{\theta}}
\newcommand{\loss}{\mathcal L}
\newcommand{\net}{\mathcal N}
\newcommand{\x}{\boldsymbol{x}}
\usepackage[hidelinks]{hyperref}
\usepackage{comment}

\begin{document}
\title{PINN-LPCDD: Physics-Informed Neural Network with Learnable and Physically Consistent Domain Decomposition for Compressible Multi-Medium Flows}

\author[Zeng W H et~al.]{Weiheng Zeng\affil{1}, Hao Wang\affil{1}, Jiaxin Zhu\affil{1}, Kun Wang\affil{1,2}, Tiegang Liu\affil{1,2}\comma*, and Chengliang Feng\affil{1}\comma\corrauth}

\address{\affilnum{1}\ LMIB and School of Mathematical Sciences, Beihang University,
          Beijing 100191, P.R. China. \\
           \affilnum{2}\ International Research Center for Mathematics and Interdisciplinary Sciences, Hangzhou International Innovation Institute of Beihang University, Hangzhou 311115, P.R. China.}
 \emails{{\tt liutg@buaa.edu.cn} (T.~Liu), {\tt charlon\_feng@buaa.edu.cn} (C.~Feng)}

\begin{abstract}
The physics-informed neural network (PINN) faces significant challenges when solving compressible multi-medium flows, including the inability to locate interface positions, poor approximation of discontinuities, and difficulty in handling high-pressure-ratio problems between media with large stiffness differences. To address these issues, we propose PINN with Learnable and Physically Consistent Domain Decomposition (PINN-LPCDD), a framework that incorporates the sharp interface method (SIM) into PINNs for solving compressible multi-medium flows. Firstly, we design separate sub-networks for each medium with learnable parameters to automatically learn interface positions, so that the domain decomposition is physically consistent, and the contact discontinuity at the interface is naturally sharp. Secondly, the Euler equations with artificial viscosity are solved to enhance robustness when handling problems involving shock waves. Lastly, the scaling invariance of the Euler equations is exploited to facilitate the solution of high pressure ratio problems. Numerical experiments on air-air, air-helium, air-water, and air-R22 problems demonstrate the acceptable accuracy of the proposed method in simulating multi-medium flows.
\end{abstract}

\pac{}

\maketitle

\section{Introduction}
\label{sec:intro}

Compressible multi-medium flows arise in a wide range of scientific and engineering applications such as inertial confinement fusion \cite{clark2019three}, underwater explosions \cite{klaseboer2005experimental}, and bubble dynamics \cite{matula1997bjerknes}. Unlike single-medium problems, where the entire flow field is governed by a single equation of state (EOS), multi-medium flows involve multiple EOS regions separated by interfaces. These interfaces are contact discontinuities across which the density, sound speed, and EOS parameters jump, while pressure and normal velocity remain continuous. When shock waves interact with these interfaces, complex wave patterns emerge. Consequently, solving multi-medium compressible flows demands not only accurate shock capturing but also robust interface tracking or capturing methods, together with a consistent treatment of the EOS discontinuity, which renders the problem considerably more challenging than its single-medium counterpart.

High-resolution numerical methods, represented by the Weighted Essentially Non-Oscillatory (WENO) scheme \cite{jiang1996efficient, friedrich1998weighted, hu1999weighted} and the Discontinuous Galerkin (DG) method \cite{reed1973triangular, cockburn1989tvb, cockburn1991runge}, have been extensively developed for shock-capturing simulations. Compared with single-medium problems, the key distinction of multi-medium problems lies in the identification and treatment of the interfaces between different media, which can be categorized into two families. The first is the diffuse interface method (DIM), whose representative technical route is to solve the five-equation model \cite{allaire2002five} and the $\gamma$-based model \cite{shyue2006efficient}; by regularizing the interface into a smooth transition layer, it can be solved directly by WENO or DG schemes without interface reconstruction. The second is the sharp interface method (SIM), whose representative technical route is to capture the interface with the front track method \cite{glimm1998three, tryggvason2001front} or level set method \cite{osher1988fronts, sethian1999level}, and to construct the interface conditions with the ghost fluid method (GFM) \cite{fedkiw1999non} and its variant \cite{liu2003ghost, liu2005ghost}.

In recent years, the emerging paradigm of artificial intelligence for PDE (AI4PDE) has garnered significant attention in computational mathematics. Current AI4PDE methodologies for multi-medium flows can be broadly classified into two distinct categories: (I) hybrid approaches that leverage deep learning to enhance traditional high-resolution numerical schemes, and (II) end-to-end frameworks that employ deep neural networks to approximate the solutions directly.

For the first category of approaches, machine learning is typically employed as a modular component to improve specific sub-tasks of traditional schemes, such as troubled-cell detection \cite{ray2018artificial, feng2020characteristic, feng2021characteristic}, numerical flux computation \cite{nguyen2022machine, morand2024deep}, and artificial viscosity selection \cite{yu2022data, bois2024optimal}. In the context of multi-medium flows, methods for accelerating accurate multi-medium Riemann solvers \cite{xu2024physics, xu2025unsupervised} and designing approximate multi-medium Riemann solvers \cite{zhang2025physics} have also been proposed.

For the second category of approaches, a predominant strategy is the physics-informed neural network (PINN) framework introduced by Raissi et al. \cite{raissi2019physics}, which embeds the governing PDEs and their initial and boundary conditions into the loss function of the network. In recent years, PINNs have been applied to multi-medium incompressible flows \cite{buhendwa2021inferring, zhu2023physics, radhakrishnan2024distributed, zhou2024self, zhai2026physics} and the compressible five-equation model \cite{liu2026data}. However, few works have applied the sharp interface method to PINN-based compressible flow solving, and such a paradigm faces the following three difficulties:

(I) The identification and separation of different media. Since different media are governed by different EOSs and separated by interfaces, a common remedy is to decompose the spatio-temporal domain into sub-domains and to deploy separate neural networks on each sub-domain, for example, cPINN/XPINN \cite{jagtap2020conservative, jagtap2020extended}. However, these two methods determine the interface positions of each sub-domain before training, whereas the evolution of interfaces in multi-medium flows needs to be learned during the training process. Moreover, physically consistent conditions should be imposed at the interfaces, rather than directly imposing continuity conditions on all variables. 

(II) The accurate capturing of shock waves. This is a long-standing challenge for PINNs applied to conservation laws, and a rich body of methods has been developed, for example, solving the Euler equations with artificial viscosity \cite{urban2025riemann, jnini2026curvature, lei2025discontinuity, wassing2025physics, li2026adaptive}, imposing the Rankine-Hugoniot jump conditions \cite{liu2024discontinuity, zeng2026scinn,ding2026adaptive}, adaptively refining collocation sampling in shock regions \cite{lu2021deepxde, mao2023physics}, and adjusting the loss weights of PDE residuals at different collocation points \cite{ferrer2024gradient, song2026physics}. Further methods can be found in a recent review \cite{abbasi2025challenges}. However, these methods still cannot fundamentally address the problem of approximating discontinuities, which remains an open problem.

(III) The data scale disparity caused by large pressure ratios. Taking the air-water Riemann problem as an example, the pressure ratio reaches $10^4{:}1$, and such a large disparity in data scales poses difficulties for neural network approximation. Existing solutions to this issue involve data preprocessing, including z-score normalization \cite{xu2025preprocessing} and logarithmic transformation \cite{peyvan2024riemann}. However, how these methods can be applied to the Euler equations requires further investigation. 

To handle the above difficulties, we propose PINN-LPCDD, a novel framework for simulating compressible multi-medium flows.  The main contributions are summarized as follows.
\begin{itemize}
  \item We propose, to the best of our knowledge, the first framework that incorporates SIM into PINNs for compressible multi-medium flows. The core is a learnable and physically consistent domain decomposition strategy: the domain interfaces carry clear physical meaning, corresponding to the material interfaces. Since the interface equations are unknown a priori, they must be learned by enforcing the physical continuity conditions across the interfaces, rather than being fixed before training as in cPINN/XPINN. Numerical experiments show that this learning strategy achieves acceptable accuracy, and on some of 2D test cases considered in this paper, its computational efficiency is close to that of traditional methods.
  \vspace{-0.25em}
  \item We design different capturing schemes for different types of discontinuities. Shocks are captured by solving the Euler equations with artificial viscosity, while the aforementioned learnable domain decomposition strategy enables the contact discontinuities at the interfaces to be captured and to remain naturally sharp. Numerical experiments validate the effectiveness of each capturing scheme.
  \vspace{-0.25em}
  \item We leverage the scaling invariance of the Euler equations to preprocess the physical variables, in order to address the issue of large pressure ratios. Numerical experiments demonstrate the necessity of data preprocessing in multi-medium flows with large stiffness differences.
\end{itemize}

\section{Preliminaries}
\label{sec:pre}

In this section, we introduce the governing equations for two-medium compressible flows and discuss the two predominant paradigms for solving the Euler equations with physics-informed neural networks.

\subsection{Euler equations for two-medium flows}
We first consider the two-dimensional (2D) Euler equations
\begin{equation}
    \big(\partial_t V+\nabla\cdot \boldsymbol F(V)-S(V)\big)\Big|_{(\x,t)}=0 ,\quad \x=(x,y)\in\R^2,\ t>0 
\label{euler_pde}
\end{equation}
where $V,\ \boldsymbol F(V)$ are defined as
\begin{equation}V=\begin{pmatrix}
    \rho\\ \rho u\\ \rho v\\ E
    \end{pmatrix},\quad 
    \boldsymbol F(V)=\begin{pmatrix}
    \rho u & \rho v\\ \rho u^2+p & \rho uv\\ \rho uv & \rho v^2+p\\ u(E+p) & v(E+p)
    \end{pmatrix}.
\label{variables}
\end{equation}
Here, $\rho$, $\boldsymbol u=(u,v)$ and $p$ represent the density, velocity and pressure, respectively; and $E$ stands for the total energy. The stiffened gas equation of state (EOS) is employed to close Eq.~\eqref{euler_pde}, so that $E$ can be related to the aforementioned variables through
\[
E = \frac{p + \Gamma B}{(\Gamma - 1)}+\frac12 \rho (u^2+v^2),
\]
where $\Gamma$ denotes the ratio of specific heats, and $B$ is the stiffness parameter of the stiffened gas equation of state. Both of them are material parameters that are constant within each medium. Consequently, Eq.~\eqref{euler_pde} can be rewritten as \begin{equation}\mathcal R(U(\x,t),\Gamma(\x,t),B(\x,t))=0, \label{euler_pde_mm} \end{equation} where $U:=(\rho,\boldsymbol u,p)^{\mathsf T}$ denotes the primitive variables, which are also the unknowns to be solved by the neural network method in this paper. Note that Eq.~\eqref{euler_pde_mm} is written differently from the single-medium problem: since $\Gamma$ and $B$ are constants everywhere, it can simply be written as $\mathcal R(U(\x,t))=0$. 

In Eq.~\eqref{euler_pde}, $S$ denotes the source term. In this paper we consider the geometric source term
\begin{equation}
S(V)=-\frac{\alpha}{x}\left(
    \rho u,\ \rho u^2,\ \rho u v,\ u(E+p)\right)^{\mathsf T}
\label{source}
\end{equation}
which accounts for the geometric effect of the computational domain: when $\alpha=0$, we have plane 2D flow, and when $\alpha=1$, it models 3D cylindrically symmetric flow.

At the material interface, the media on both sides should satisfy the continuity conditions. Let $\boldsymbol n=(n_x,n_y)$ denote the unit normal vector to the interface. The jump in the normal velocity across the interface vanishes, i.e., the normal velocity is continuous across the interface:
\begin{equation}
[\boldsymbol u\cdot \boldsymbol n] = [u n_x + v n_y] = 0 ; \quad[p] = 0. 
\label{interface_condition}
\end{equation}

Note that when the $y$-dependence and the velocity component $v$ are omitted, Eq.~\eqref{euler_pde} reduces to the 1D Euler equations. Then in Eq.~\eqref{source}, $\alpha=0,1,2$ correspond to 1D plane flow, 2D cylindrically symmetric flow, and 3D spherically symmetric flow, respectively.

\subsection{Paradigms for solving Euler equations with PINNs}
\label{sec:paradigms}

The Euler equations are time-dependent equations. Based on the treatment of time, PINN methods can be roughly divided into the following two types.

\paragraph{Paradigm 1: full space--time domain regression.}
This approach treats time as a dimension equivalent to space. It seeks a neural network $\net_{\sita}$ that approximates the primitive variables $U(\x,t)$ over the entire space--time domain:
\begin{equation}
\bigl\|U(\x,t) - \net_{\sita}(\x,t)\bigr\| \to 0.
\label{paradigm1}
\end{equation}
The PDE residual is evaluated by automatic differentiation of the network output, and the loss function penalizes violations of the governing equations. The underlying philosophy is that minimizing the PDE residual (compatibility condition) drives the network toward the true solution. The well-known domain decomposition method cPINN \cite{jagtap2020conservative} belongs to this category.

\paragraph{Paradigm 2: time-stepping operator learning.}
This approach learns a time-marching scheme that maps the solution from one time level to the next. It seeks a neural network $\net_{\sita}$ such that
\begin{equation}
\bigl\|U^{n+1}(\x) - \net_{\sita}(U^n(\x))\bigr\| \to 0.
\label{paradigm2a}
\end{equation}
or, in an explicit Euler-type formulation,
\begin{equation}
\bigl\|U^{n+1}(\x) - U^n(\x) - \Delta t\,\net_{\sita}(U^n(\x))\bigr\| \to 0.
\label{paradigm2b}
\end{equation}
Here the PDE is discretized by a numerical scheme and the loss function measures the deviation from the discrete conservation law. The essence of this paradigm is to learn a data-driven time-stepping rule. The five-equation PINN model for compressible multi-medium flows \cite{liu2026data} falls into this category.

This paper adopts the first paradigm. We aim to improve the PINN formulation so that the three difficulties outlined in the introduction are alleviated, and the neural network solution converges as closely as possible to the exact solution.

\section{Methodology}
\label{sec:method}

In this section, we present the neural-network-based framework for solving the initial-boundary value problem of the Euler equations with two media:
\begin{equation}
    \begin{cases}
        \mathcal R(U(\x,t),\Gamma(\x,t),B(\x,t))=0 ,& (\x,t)\in \Omega\times[0,T]
        \\
        (U, \Gamma, B)|_{t=0}=(U_Q(\x), \Gamma_Q, B_Q),\ & \x\in \Omega_{0,Q}, Q=L/R
        \\
        \mathcal B(U(\x,t))=0, & (\x,t)\in \partial \Omega\times[0,T]
    \end{cases}
\label{problem_definition}
\end{equation}
where $\mathcal R$ is defined in Eq.~\eqref{euler_pde}, $\Omega \subseteq \R^n (n=1,2)$ is the spatial domain, $\partial \Omega$ denotes its boundary, and $T \in \R_+$ is the final time.  Moreover, $\Omega_{0,Q}$ denotes the region occupied by medium $Q$ ($Q=L,R$), and $\Omega$ is the disjoint union of $\Omega_{0,L}$ and $\Omega_{0,R}$.  The boundary condition $\mathcal B(U)=0$ is specified according to the physical setup, and we consider the following three types in this paper:
\begin{itemize}
    \item \emph{Outflow condition}: no restriction is imposed on the values of $U$ at the boundary;
    \item \emph{Dirichlet condition}: $U(\x,t)=U_B(\x,t)$ for $(\x,t)\in \partial \Omega\times[0,T]$, where $U_B$ is the prescribed boundary function;
    \item \emph{Reflection (symmetry) condition}: $\partial_{\boldsymbol n}\rho=\partial_{\boldsymbol n}p=0$ and $\boldsymbol u\cdot \boldsymbol n=0$ on $\partial\Omega$, where $\boldsymbol n$ denotes the outward unit normal vector to $\partial\Omega$.
\end{itemize}

\subsection{Data preprocessing}
\label{sec:preprocessing}

When simulating multi-medium flows with PINNs, the primitive variables $(\rho, \boldsymbol u, p)$ can vary over several orders of magnitude. For example, the pressure ratio reaches $10^4{:}1$ in air-water Riemann problems. Without proper preprocessing, such large variations lead to ill-conditioned loss landscapes and hinder convergence. 

In this subsection, we employ a scaling transformation strategy to address this issue. The Euler equations are invariant under a two-parameter scaling group, which provides a natural mechanism for normalizing the variables. 

\begin{proposition}
\label{prop:scaling}
Let $(\rho,\boldsymbol u,p)$ be a solution of Eq.~\eqref{euler_pde}. For any $\lambda, \mu>0$, set $\tilde t=\lambda t$ and define the rescaled fields
\begin{equation}
\tilde\rho(\x,\tilde t)=\frac{1}{\mu}\rho(\x,t),\quad
\tilde{\boldsymbol u}(\x,\tilde t)=\frac{1}{\lambda}\boldsymbol u(\x,t),\quad
\tilde p(\x,\tilde t)=\frac{1}{\mu\lambda^2}p(\x,t),\quad
\tilde B(\x,\tilde t)=\frac{1}{\mu\lambda^2}B(\x,t).
\label{scaled_fields}
\end{equation}
Then the total energy satisfies $\tilde E=\frac{1}{\mu\lambda^2}E$, and the new variables also satisfy Eq.~\eqref{euler_pde}.
\end{proposition}

Consequently, the computational procedure consists of the following three steps: 
    (I) transform $(t,\rho,\boldsymbol u,p,B)$ to $(\tilde t,\tilde\rho,\tilde{\boldsymbol u},\tilde p,\tilde B)$ by Eq.~\eqref{scaled_fields};
    (II) solve Eq.~\eqref{euler_pde} with the new variables on $\Omega\times[0,\lambda T]$;
    (III) transform all the variables back to the physical ones.

\subsection{Interface parameterization}
\label{sec:rls}

When the SIM is applied to PINNs, the temporal evolution of the interface needs to be learned explicitly. A straightforward idea is to directly construct a neural network surrogate for the level set function $\Phi(\x,t)$ of the interface. However, this greatly increases the learning cost of PINNs. To reduce this cost, we parameterize the interface, so that the network only needs to learn the additional parameters representing the interface. This subsection focuses on this parameterization method.

We first discretize the time interval $[0,T]$ uniformly by
\begin{equation}
0=t_0<t_1<\cdots<t_K=T,\qquad t_k=k\,\Delta t,\qquad \Delta t=\frac{T}{K},
\label{rls_time}
\end{equation}
and assume that the interface equation is the zero level set of the function $\Phi(\x,t)$.

\paragraph{1D cases.} The interface equation can always be written as $\Phi(x,t)=x-s(t)=0$. Therefore, following the temporal discretization described above, we can discretize the interface $x=s(t)$ into $\{(x_k,t_k)\}_{k=0}^{K}$, and assume that adjacent time steps satisfy the following relation
\begin{equation}
x_k=x_{k-1}+\Delta t\,u_{*,k},\quad k=1,\dots,K,
\label{interface1D}
\end{equation}
with $x_0$ being the initial interface position. Then the interface is parameterized by $\{u_{*,k}\}_{k=1}^K$ and learned by the neural network. If the interface velocity is constant, then the interface can be represented by a single parameter $u_*=u_{*,1}=\cdots=u_{*,K}$.

\paragraph{2D cases.} We suppose the interface $\Phi(\cdot, t_k)=0$ is a curve with a start point $\x_{k,0}$ and an end point $\x_{k,J}, J\in \mathbb N_+$. The points $\{\x_{k,j}\}_{j=1}^J$ are sampled successively between the start and end points, and their temporal evolution is tracked via the following formula
\begin{equation}
\x_{k,j}=\x_{k-1,j}+\Delta t\,\boldsymbol{u}_{*,k,j},\quad k=1,\dots,K;\ j=0,\dots,J,
\label{interface2D}
\end{equation}
with $\{\x_{0,j}\}_{j=0}^J$ the sampling points of the initial interface. Then the interface is parameterized by $\{\boldsymbol{u}_{*,k,j}\}_{k=1,j=0}^{k=K,j=J}$ and learned by the neural network. 

At this point, the unit normal vector $\boldsymbol n_{k,j}$ at the interface point $\x_{k,j}=(x_{k,j},y_{k,j})$ can be approximated by the normal vector of the line segment connecting $\x_{k,j-1}$ and $\x_{k,j+1}$:
\begin{equation}
\boldsymbol n_{k,j}=\frac{(y_{k,j+1}-y_{k,j-1},-x_{k,j+1}+x_{k,j-1})}{\sqrt{(y_{k,j+1}-y_{k,j-1})^2+(x_{k,j+1}-x_{k,j-1})^2}},\quad k=1,\dots,K;\ j=1,\dots,J-1.
\label{interface_normal}
\end{equation}
If the interface at a fixed time is a closed curve, i.e., $\x_{k,0}=\x_{k,J}$, then we set $\x_{k,\pm1}=\x_{k,J\pm1}$, so that the normal vectors at the first and last points can also be determined by the above formula.

For notational convenience, in the following we consistently use the notation of Eq.~\eqref{interface2D} to describe interface-related problems in both the 1D and 2D cases. When computing 1D problems, $J\equiv 0$, meaning that there is only one interface point at each time instant; and $\boldsymbol n_{k,j}\equiv 1$ represents the outward normal direction at the interface point. All the interface parameters are initialized to zero before training.

\begin{remark}
The above parameterization follows the spirit of the classical front-tracking method \cite{glimm1998three, tryggvason2001front}: the interface is represented explicitly by a set of marker points that are advanced in a Lagrangian fashion, and the interface geometry, e.g., the unit normal vectors in Eq.~\eqref{interface_normal}, is reconstructed from the neighboring markers. The key difference lies in the determination of the marker velocity. In classical front tracking, the markers are advected by the fluid velocity interpolated from the underlying flow field, whereas in our method the interface velocity is parameterized by a set of learnable parameters, which are determined during training by enforcing the physical continuity conditions (Eq.~\eqref{interface_condition}) across the interface. The specific enforcement of these conditions is detailed in Subsection~\ref{sec:loss}.
\end{remark}

\subsection{Network architecture}
\label{sec:arch}

The network architecture is designed with the discontinuities in compressible multi-medium flows in mind, addressing the following two aspects:
\begin{itemize}
    \item \emph{Shock representation}: shocks are intrinsically high-frequency and large-gradient features, whereas neural networks suffer from the spectral bias \cite{rahaman2019spectral}, i.e., high-frequency components are learned much more slowly than low-frequency ones. Two complementary mechanisms are therefore incorporated: a Fourier feature embedding \cite{tancik2020fourier}, which encodes the network input with high-frequency functions to overcome the spectral bias, and an RBF output layer \cite{xiao2022hybrid, li2026adaptive}, whose localized Gaussian kernels exert strong local responses in the vicinity of large-gradient regions and thereby enable high-precision local refinement of shock waves.
    \item \emph{Sharp interface representation}: SIM requires the solution to be genuinely discontinuous across the material interface, specifically in the density and the velocity component tangential to the interface, whereas a single network composed of continuous activation functions is itself continuous and can only approximate a discontinuity with a steep but smooth transition layer. We therefore assign one network to each medium, so that the jump in the solution across the interface is preserved by construction, with the interface position determined by the learnable parameters introduced in the preceding subsection.
\end{itemize}

Below, we elaborate on the three components, i.e., the Fourier feature embedding, the fully-connected sub-networks, and the RBF output layer, respectively.

\paragraph{Fourier feature embedding.}
To alleviate the spectral bias discussed above, an additional non-trainable layer is introduced, which encodes the input vector of the neural network using a set of sinusoidal functions with randomly sampled frequencies:
\begin{equation}
\hat{\boldsymbol u}_{\mathrm{fe}}(\boldsymbol r):\R^{D_0}\rightarrow\R^{2D_{\mathrm{fe}}},
\label{fe_map}
\end{equation}
\begin{equation}
\begin{pmatrix}
\hat u_{{\mathrm{fe}},i}\\
\hat u_{{\mathrm{fe}},i+1}\\
\vdots
\end{pmatrix}
=
\begin{pmatrix}
\sum_{j=1}^{D_0}\sin(\phi_{i,j}\,r_j)\\
\sum_{j=1}^{D_0}\cos(\phi_{i,j}\,r_j)\\
\vdots
\end{pmatrix},
\qquad i=2k-1,\quad k=1,\dots,D_{\mathrm{fe}},
\label{fe_embedding}
\end{equation}
where $\boldsymbol r:=(\x,t)\in\R^{D_0}$ is the input vector ($D_0=n+1$ for the n-dimensional cases), and $\phi_{i,j}$ are the randomly sampled frequencies from a normal distribution $\mathcal N(0,1)$. Importantly, the Fourier embedding is shared by the two sub-networks $Q=L,R$, i.e., the frequency weights $\phi_{i,j}$ are common to both sub-networks, so that the two networks operate on the same encoded feature space, while the weights of the fully-connected layers remain independent for each sub-network.

\paragraph{Fully-connected sub-networks.}
Following the parameterization in Subsection~\ref{sec:rls}, the spatio-temporal domain $\Omega \times[0,T]$ is decomposed into two sub-domains separated by the material interface $\Phi(\x,t)=0$:
\begin{equation}
\begin{aligned}
\Omega_L &:=\{(\x,t)\mid \Phi(\x,t) < 0,\; \x\in\Omega,\; 0\le t\le T\},\\
\Omega_R &:=\{(\x,t)\mid \Phi(\x,t) > 0,\; \x\in\Omega,\; 0\le t\le T\},
\label{domain_decomp}
\end{aligned}
\end{equation}
where $\Phi(\x,t)<0$ and $\Phi(\x,t)>0$ correspond to medium $L$ and medium $R$, respectively. On each sub-domain $\Omega_Q$ ($Q=L,R$), we deploy a separate network $U_{\sita, Q}$ to approximate the primitive variables $U:=(\rho,\boldsymbol u,p)^\top$.

The architecture of each sub-network follows the standard PINN design \cite{raissi2019physics}. Consider a backbone network with $M-1$ hidden layers, the $\ell$-th layer containing $n_\ell$ neurons. Taking the shared Fourier embedding as the input, the computation flow of the backbone of sub-network $Q$ can be expressed as:
\begin{equation}
\begin{aligned}
h_Q^0(\x,t) &= \hat{\boldsymbol u}_{\mathrm{fe}}(\boldsymbol r) \in \R^{2D_{\mathrm{fe}}},\\[2mm]
h_Q^m(\x,t) &= \sigma\big(\boldsymbol{W}_Q^m h_Q^{m-1}(\x,t) + \boldsymbol{b}_Q^m\big) \in \R^{n_m},\qquad m = 1,\dots,M-1,
\label{ffn_arch}
\end{aligned}
\end{equation}
where $\boldsymbol{W}_Q^m \in \R^{n_m \times n_{m-1}}$ and $\boldsymbol{b}_Q^m \in \R^{n_m}$ are the weight matrix and bias vector of the $m$-th layer in sub-network $Q$, with $n_0=2D_{\mathrm{fe}}$; $\sigma(\cdot)$ is the activation function (e.g., $\tanh$ in this paper) that introduces nonlinearity, and the weight matrices and bias vectors of the fully-connected layers are initialized by the Xavier scheme \cite{glorot2010understanding}. Different from the standard feedforward network (FFN), which produces the output with a linear readout layer, the network output is computed from the last hidden-layer feature $\boldsymbol z:=h_Q^{M-1}(\x, t)$ by an RBF output layer, which is described next.

\paragraph{RBF output layer.}
To further enhance the capacity of the sub-networks for local structure fitting, an RBF layer is introduced into the feature space output by the backbone network \cite{xiao2022hybrid, li2026adaptive}. The motivation is that the fully-connected layers excel at capturing the global features of the solution, whereas the RBF layer specializes in the local approximation of the solution. The activation functions of this layer consist of $K_{\mathrm{rbf}}$ Gaussian kernel functions, and the output of the $j$-th neuron is defined as
\begin{equation}
\psi_j(\boldsymbol z) = \exp\Big( -\frac{\|\boldsymbol z - \boldsymbol c_j\|^2}{2\varsigma_j^2} \Big),\qquad j = 1,\dots,K_{\mathrm{rbf}},
\label{rbf_kernel}
\end{equation}
where $\|\boldsymbol z - \boldsymbol c_j\|^2 = \boldsymbol z^{\top}\boldsymbol z + \boldsymbol c_j^{\top}\boldsymbol c_j - 2\boldsymbol z^{\top}\boldsymbol c_j$ is the squared Euclidean distance between the feature vector $\boldsymbol z = h_Q^{M-1}(\x,t) \in \R^{n_{M-1}}$ and the center point $\boldsymbol c_j \in \R^{n_{M-1}}$, and $\varsigma_j > 0$ is the corresponding scale parameter. The prediction of sub-network $Q$ is then given by the linear output layer
\begin{equation}
U_{\sita, Q}(\x,t) = \boldsymbol{W}_Q^{\mathrm{out}}\, \boldsymbol{\psi}\big(\boldsymbol z\big) + \boldsymbol{b}_Q^{\mathrm{out}} \in \R^{n+2},
\label{rbf_output}
\end{equation}
where $\boldsymbol{\psi}(\boldsymbol z) := \big(\psi_1(\boldsymbol z), \dots, \psi_{K_{\mathrm{rbf}}}(\boldsymbol z)\big)^{\top} \in \R^{K_{\mathrm{rbf}}}$ stacks the outputs of the RBF neurons, and $\boldsymbol{W}_Q^{\mathrm{out}} \in \R^{(n+2)\times K_{\mathrm{rbf}}}$ and $\boldsymbol{b}_Q^{\mathrm{out}} \in \R^{n+2}$ are the trainable linear transformation parameters of the output layer; the output dimension $n+2$ corresponds to the primitive variables $(\rho,\boldsymbol u,p)^{\top}$ for the $n$-dimensional cases. 

In practical implementation, each RBF center is drawn component-wise from the normal distribution $\mathcal N(0,1)$ and fixed during the training phase; the scale parameters are randomly generated from the uniform distribution $\mathcal U(0.5,2.0)$ and incorporated into the joint training. The two sub-networks possess their own independent RBF output layers.

\subsection{Loss function}
\label{sec:loss}
The total loss function comprises two parts: the sub-network losses for each sub-domain and the interface continuity loss, so that both the interface parameters in Subsection~\ref{sec:rls} and the network parameters in Subsection~\ref{sec:arch} can be optimized.

\paragraph{Sub-network loss.}
For each sub-domain $\Omega_Q$ ($Q=L,R$), the governing equations are the Euler equations supplemented with an artificial viscosity term:
\begin{equation}
\partial_t V + \nabla\cdot \boldsymbol F(V) - S(V) = \nu\,\nabla^2 V,
\label{euler_viscosity}
\end{equation}
where $V$, $\boldsymbol F(V)$ and $S(V)$ are defined in Eq.~\eqref{variables}, and $\nu>0$ is a constant artificial viscosity coefficient. Since the sub-network outputs the primitive variables, we denote its prediction by $U_{\sita, Q}:=(\rho_{\sita, Q},\boldsymbol u_{\sita, Q},p_{\sita, Q})^\top$ and the corresponding conserved variables by $V_{\sita, Q}:=(\rho_{\sita, Q},\rho_{\sita, Q}\boldsymbol u_{\sita, Q},E_{\sita, Q})^\top$, and the PDE residual with artificial viscosity is then
\begin{equation}
\mathcal R_Q^\nu(U_{\sita, Q}, \Gamma_Q, B_Q) := \partial_t V_{\sita, Q} + \nabla\cdot \boldsymbol F(V_{\sita, Q}) - S(V_{\sita, Q}) - \nu\,\nabla^2 V_{\sita, Q}.
\label{pde_residual}
\end{equation}
Note that in the numerical implementation, the residuals are computed with the explicit component-wise forms provided in Appendix~\ref{app:pde}.

Following the standard PINN formulation, the loss function for sub-network $Q$ consists of three terms:
\begin{align}
\loss_{\mathrm{GOV},Q}(\mathcal R^\nu(U, \Gamma, B)) &:= \frac{1}{|P_N^Q|} \sum_{(\x_j,t_j) \in P_N^Q} \big\| \mathcal R_Q^\nu(U_{\sita, Q}, \Gamma_Q, B_Q)(\x_j,t_j) \big\|^2,\label{loss_gov}\\
\loss_{\mathrm{IC},Q} (U) &:= \frac{1}{|P_I^Q|} \sum_{\x_j \in P_I^Q} \big\| U_{\sita, Q}(\x_j,0) - U_0(\x_j) \big\|^2,\label{loss_ic}\\
\loss_{\mathrm{BC},Q} (U)&:= \frac{1}{|P_B^Q|} \sum_{(\x_j,t_j) \in P_B^Q} \big\| \mathcal B(U_{\sita, Q}(\x_j,t_j)) \|^2 ,\label{loss_bc}
\end{align}
where $P_N^Q \subset \mathrm{int}\ \Omega_Q$, $P_I^Q \subset \Omega_Q \cap \{t=0\}$, and $P_B^Q \subset \Omega_Q \cap \partial \Omega $ are collocation point sets for the governing equation, initial condition, and boundary condition, respectively. 

\paragraph{Interface continuity loss.}
The losses in Eqs.~\eqref{loss_gov}--\eqref{loss_bc} are designed separately for the two sub-networks, each restricted to its own sub-domain. How the two neural networks interact with each other, and how the interface parameters in Subsection~\ref{sec:rls} participate in the training simultaneously, rely on the interface continuity conditions.

We first consider the interaction through the pressure $p$. According to the continuity condition, the pressures on the two sides of the interface should be equal, which motivates the following loss:
\begin{equation}
\loss_{\mathrm{IF},p} := \frac{1}{|P_{\mathrm{IF}}|} \sum_{(\x_{k,j},t_k)\in P_{\mathrm{IF}}} \big( p_{\sita, L}(\x_{k,j},t_k) - p_{\sita, R}(\x_{k,j},t_k) \big)^2,
\label{loss_if_p}
\end{equation}
where $P_{\mathrm{IF}}:=\{(\x_{k,j},t_k)\mid k=1,\dots,K,\ j=1,\dots,J\}$ is the set of interface collocation points.

The continuity condition for the velocity $\boldsymbol u$ in the normal direction is analogous to that of $p$. However, its loss is designed slightly differently, since the collaborative training of the interface parameters must be taken into account. To this end, this term involves the velocities predicted by the two sub-networks at the interface points as well as the interface velocity parameters corresponding to these points, and requires all of them to be equal along the normal direction:
\begin{equation}
\loss_{\mathrm{IF},\boldsymbol u} := \frac{1}{|P_{\mathrm{IF}}|} \sum_{(\x_{k,j},t_k)\in P_{\mathrm{IF}}} \sum_{Q=L}^R \big( (\boldsymbol u_{\sita, Q}(\x_{k,j},t_k) - \boldsymbol u_{*,k,j})\cdot\boldsymbol n_{k,j} \big)^2,
\label{loss_if_u}
\end{equation}
where $\boldsymbol n_{k,j}$ is the unit normal vector at the interface point $\x_{k,j}$ given by Eq.~\eqref{interface_normal}. 

\paragraph{Total loss.}
The overall loss function is then a weighted sum of all the above terms:
\begin{equation}
\loss = \sum_{Q=L,R} \big( w_{\mathrm{GOV}} \loss_{\mathrm{GOV},Q} + w_{\mathrm{IC}} \loss_{\mathrm{IC},Q} + w_{\mathrm{BC}} \loss_{\mathrm{BC},Q} \big) + w_{\mathrm{IF}} \big( \loss_{\mathrm{IF},p} + \loss_{\mathrm{IF},\boldsymbol u} \big),
\label{total_loss}
\end{equation}
where $w_{\mathrm{GOV}}, w_{\mathrm{IC}}, w_{\mathrm{BC}}, w_{\mathrm{IF}}$ are weighting coefficients. The network parameters and the interface parameters are jointly optimized by minimizing $\loss$.

\begin{figure}[H]
\centering
\includegraphics[width=\textwidth]{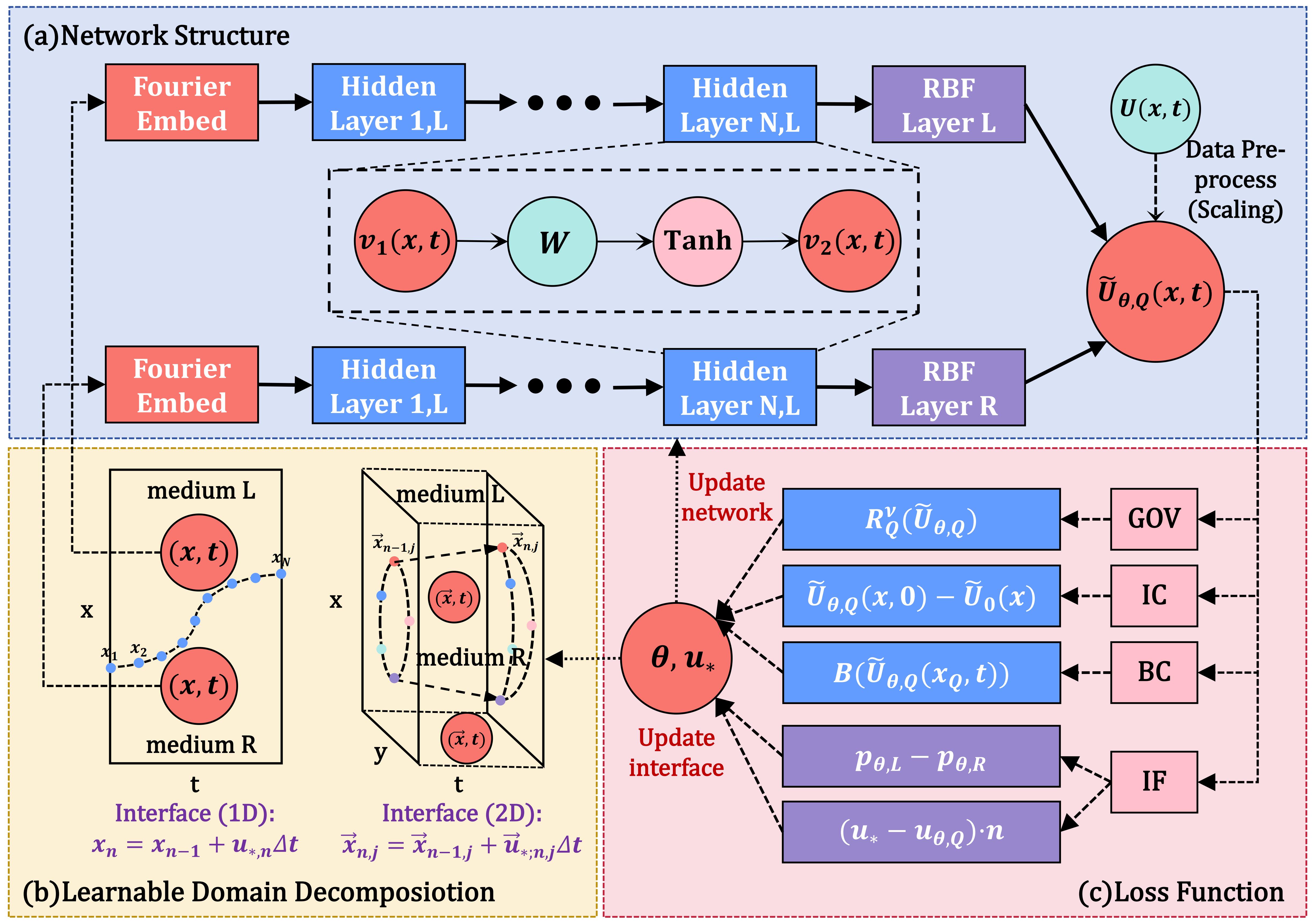}
\caption{Workflow of PINN-LPCDD for simulating compressible multi-medium flows.}
\label{fig:workflow}
\end{figure}

At the end of this section, we summarize the overall workflow of PINN-LPCDD in Fig.~\ref{fig:workflow}, which consists of three main components: (a) the network architecture, comprising a Fourier feature embedding for eliminating the spectral bias, fully-connected layers for capturing global characteristics of the solution, and an RBF layer for local refinement; (b) the learnable region partition, which learns the velocities of the discrete interface points; and (c) the loss function, consisting of the PDE loss and the initial/boundary losses for optimizing each sub-network, together with the interface continuity loss for the interaction between the two sub-networks and the interface parameters.

\section{Numerical Experiment}
\label{sec:experiment}
In this section, we conduct numerical studies on a series of problems to demonstrate the effectiveness and flexibility of PINN-LPCDD.
\begin{itemize}
    \item \textbf{Data Preparations: }We generate points $\{t_j, \boldsymbol x_j\}_{j=1}^{N_t\times N_x}$ as the input to PINN-LPCDD through a uniform partition of the computational domain. In the training stage, $(N_t, N_x)$ is set to $(64, 512)$ for 1D problems, and $(N_t, N_x, N_y)$ is set to $(16, 128, 64)$ for 2D problems. In addition, we prepare the exact or reference solutions for each test case. For source-free Riemann problems, the exact solutions are computed directly using an exact Riemann solver \cite{toro2013riemann}; otherwise, the reference solutions are obtained using the HODG platform \cite{he2023hodg}, which has been customized for multi-medium problems \cite{wang2026conservation}. All the data is prepared on an Intel(R) Core(TM) i7-14700KF CPU with 16 GB of memory. Note that our method is data-free: the exact or reference solutions are only used to evaluate the accuracy of the predicted solution, and are never used in the loss function or the training process.

    \item \textbf{Default settings: } We implement PINN-LPCDD with each sub-network being a Fourier feature embedding with $D_{\mathrm{fe}}=20$, followed by a 7-hidden-layer architecture with 20 neurons per layer and an RBF output layer with $K_{\mathrm{rbf}}=64$ Gaussian kernels. Each sub-network is first trained with the Adam algorithm for 3000 epochs and then with the SOAP algorithm \cite{vyas2025soap} for 2000 additional epochs, with a learning rate of $10^{-3}$ and a reduction factor of 0.98 every 100 epochs.  The loss function in Eq.~\eqref{total_loss} incorporates weighted components with $(w_\mathrm{GOV}, w_\mathrm{IC}, w_{\mathrm{BC}}, w_{\mathrm{IF}})=(1,20,10,10)$. We also set $\mu = \lambda = 1$ in Eq.~\eqref{scaled_fields}, and $\alpha=0$ in Eq.~\eqref{euler_pde}, and take $\nu = 2 \times 10^{-4}$ if artificial viscosity is employed. The PINN-LPCDD model is constructed with the PyTorch framework and run on an NVIDIA GeForce RTX 4070 SUPER card with 12GB of memory.
    
    \item \textbf{Model Indicators: } We evaluate the mean squared error (MSE) and the relative $L^2$ error (RE) of the variables at the terminal time. Given the exact values $\{y_i\}_{i=1}^N$ and predicted values $\{\hat{y}_i\}_{i=1}^N$, the RE is defined as
    \begin{equation}
        E_{\mathrm{rel}} = \frac{\sqrt{\sum_{i=1}^{N} \left( \hat{y}_i - y_i \right)^2}}{\sqrt{\sum_{i=1}^{N} y_i^2}},
        \label{eq:rel_l2}
    \end{equation}
    Specifically, in the following tables, MSE ($\rho/u/v/p$) denote the MSEs of the corresponding primitive variables, respectively; MSE (avg) and RE (avg) denote the averages of the MSE and RE over all the primitive variables, respectively.

\end{itemize}

\subsection{Air-air Riemann problems}
\label{sec:air_air}

In this subsection, we demonstrate the capability of PINN-LPCDD in solving single-medium problems. Here, PINN-LPCDD treats the contact discontinuity as the interface, the left and right sides as two distinct media, and learns the propagation speed of the contact discontinuity. We also compare PINN-LPCDD with two representative shock-handling approaches, namely PINN-WE, which equips the loss with the Rankine-Hugoniot relation \cite{liu2024discontinuity}, and PINN-AV, which introduces artificial viscosity. In addition, the vanilla PINN without any discontinuity-handling technique is included as a baseline.

\subsubsection{Sod problem}

This test case examines the capability of PINN-based methods to approximate rarefaction waves, weak shock waves, and weak contact discontinuities \cite{sod1978survey}. The initial conditions are set to be

\begin{equation}
\mathrm{(1A)}\quad  
(\rho,u,p,\Gamma,B)=\begin{cases}
(1.000,\ 0.0,\ 1.000,\ 1.4,\ 0.0), & x<0.5; \\
(0.125,\ 0.0,\ 0.100,\ 1.4,\ 0.0), & x>0.5,
\end{cases}
\end{equation}
and we solve this problem on $(x,t)\in[0,1]\times[0,0.2]$.

To investigate the performance of PINN-LPCDD, we compare it with PINN, PINN-AV, and PINN-WE. We implement PINN and PINN-AV on the same computational platform as PINN-LPCDD, while the results of PINN-WE \cite{liu2024discontinuity} are directly extracted from its source code repository. The difference between PINN and PINN-AV lies only in the artificial viscosity coefficient, which is $\nu=0$ for PINN and $\nu=2\times10^{-4}$ for PINN-AV. The differences between PINN-AV and PINN-LPCDD are: (1) PINN-AV uses a single network with 30 neurons per layer, ensuring that its parameter count slightly exceeds that of PINN-LPCDD; (2) PINN-AV does not need to learn the interface velocity and compute the interface loss; (3) PINN-AV does not adopt the Fourier feature embedding, whereas PINN-LPCDD does. The remaining parameter settings are identical to those of PINN-LPCDD. We evaluate the solution at the terminal time on a uniform grid with $N_x = 100$, the same as in PINN-WE. 

\begin{table}[H]
\caption{Comparison among PINN-based models in 1D Sod problem (test case (1A)).}\label{tab_compare_1A}
\small
\centering
\begin{tabular}{|c|cccc|c|}
\hline
Model & 
MSE ($\rho$) &
MSE ($u$) & 
MSE ($p$) &
MSE (avg) &
RE (avg) 
\\ \hline

PINN-LPCDD	& 1.538e-05	& 4.779e-05	& 2.219e-05	& 2.845e-05	& \bf{8.289e-03}	\\
PINN-AV	& 7.682e-04	& 1.183e-02	& 7.437e-04	& 4.447e-03	& 8.791e-02	\\
PINN-WE	& 5.105e-05	& 8.068e-05	& 2.343e-05	& 5.172e-05	& 1.113e-02	\\
PINN	& 2.848e-03	& 1.856e-03	& 2.299e-04	& 1.645e-03	& 5.887e-02	\\
\hline

\end{tabular}
\end{table}

\begin{figure}[H]
    \centering
    \includegraphics[width=\linewidth]{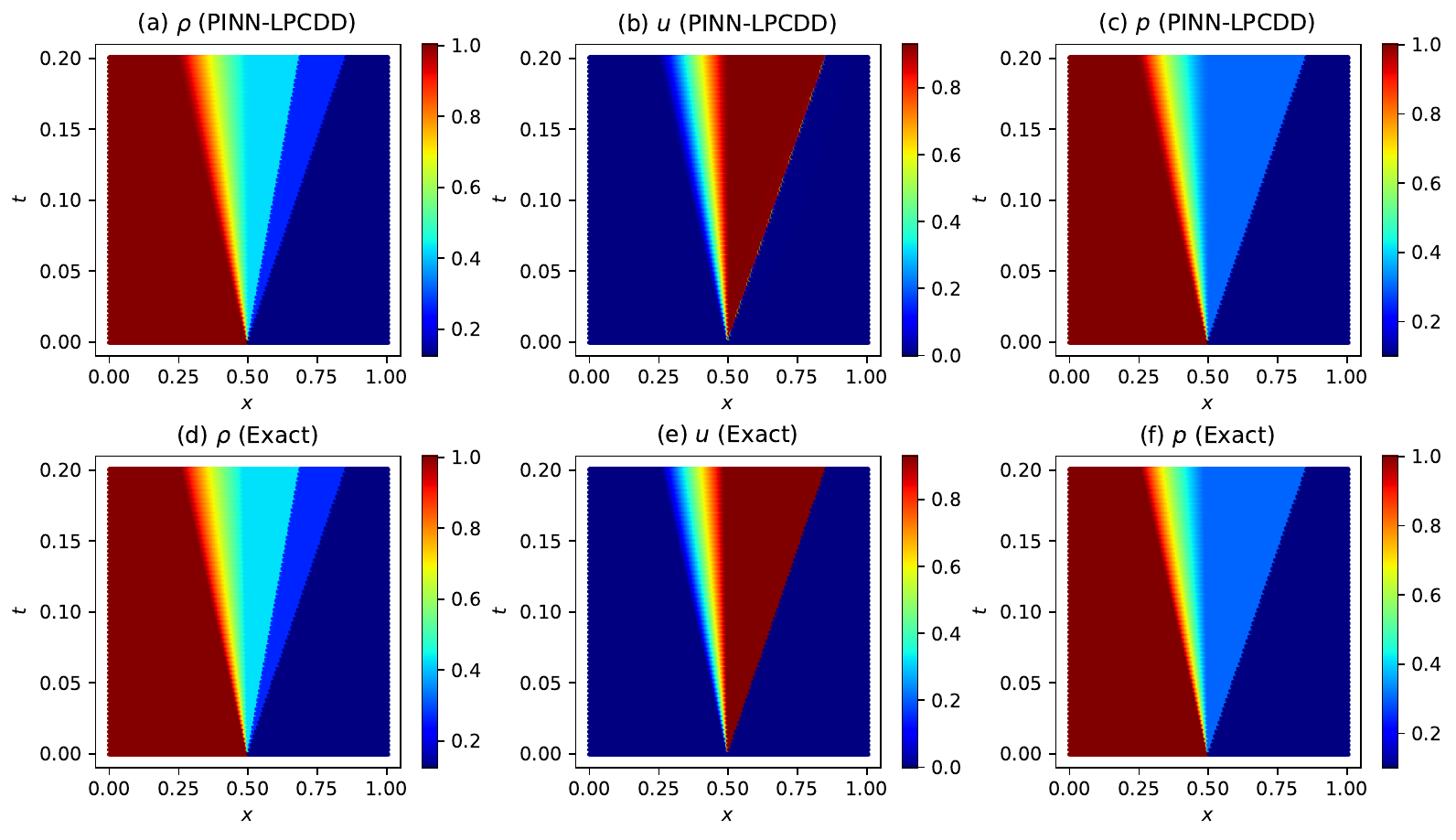}
    \includegraphics[width=\linewidth]{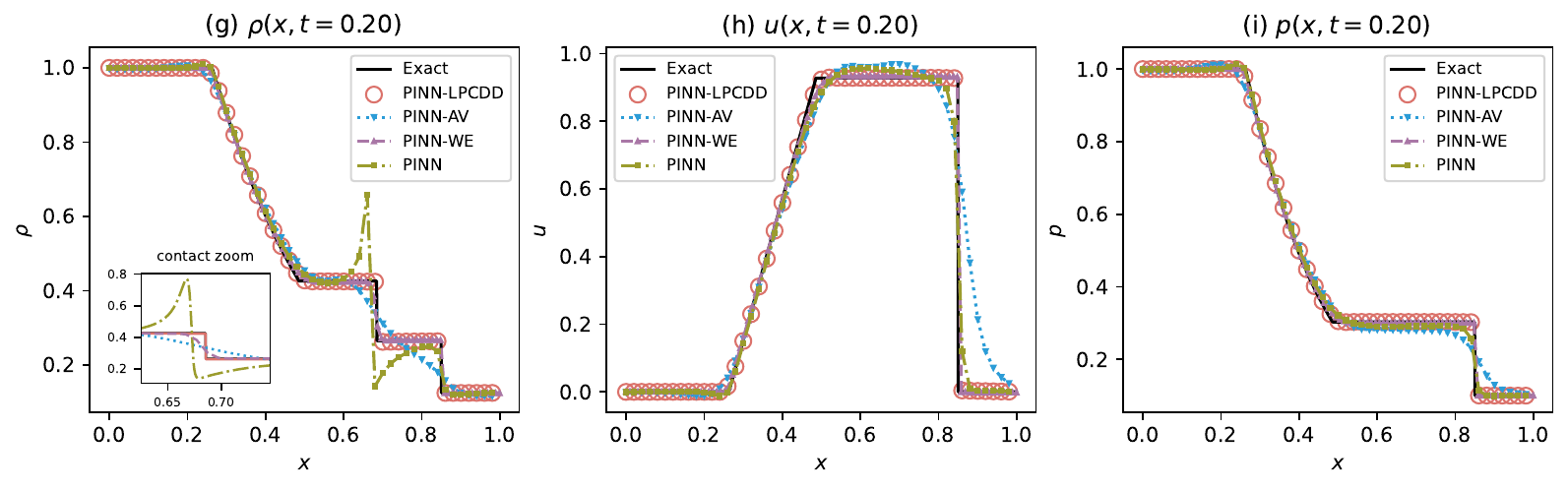}
    \caption{Results of 1D Sod problem (test case (1A)) with PINN-based methods.}
    \label{fig_compare_1A}
\end{figure}

It can be observed from Fig.~\ref{fig_compare_1A} that PINN-LPCDD captures the contact discontinuity sharply without any smearing compared with the other three models, and without the artificial viscosity, the vanilla PINN struggles to approximate the discontinuity. The results in Table~\ref{tab_compare_1A} further show that PINN-LPCDD achieves the smallest errors among all the methods.

\subsubsection{Lax problem}

This test case examines the capability of PINN-based methods to approximate rarefaction waves, strong shock waves, and strong contact discontinuities \cite{lax1954weak}. The initial conditions are set to be

\begin{equation}
\mathrm{(1B)}\quad  
(\rho,u,p,\Gamma,B)=\begin{cases}
(0.445,\ 0.698,\ 3.528,\ 1.4,\ 0.0), & x<0.5; \\
(0.500,\ 0.000,\ 0.571,\ 1.4,\ 0.0), & x>0.5,
\end{cases}
\end{equation}
and we solve this problem on $(x,t)\in[0,1]\times[0,0.13]$.

 \begin{figure}[H]
    \centering
    \includegraphics[width=\linewidth]{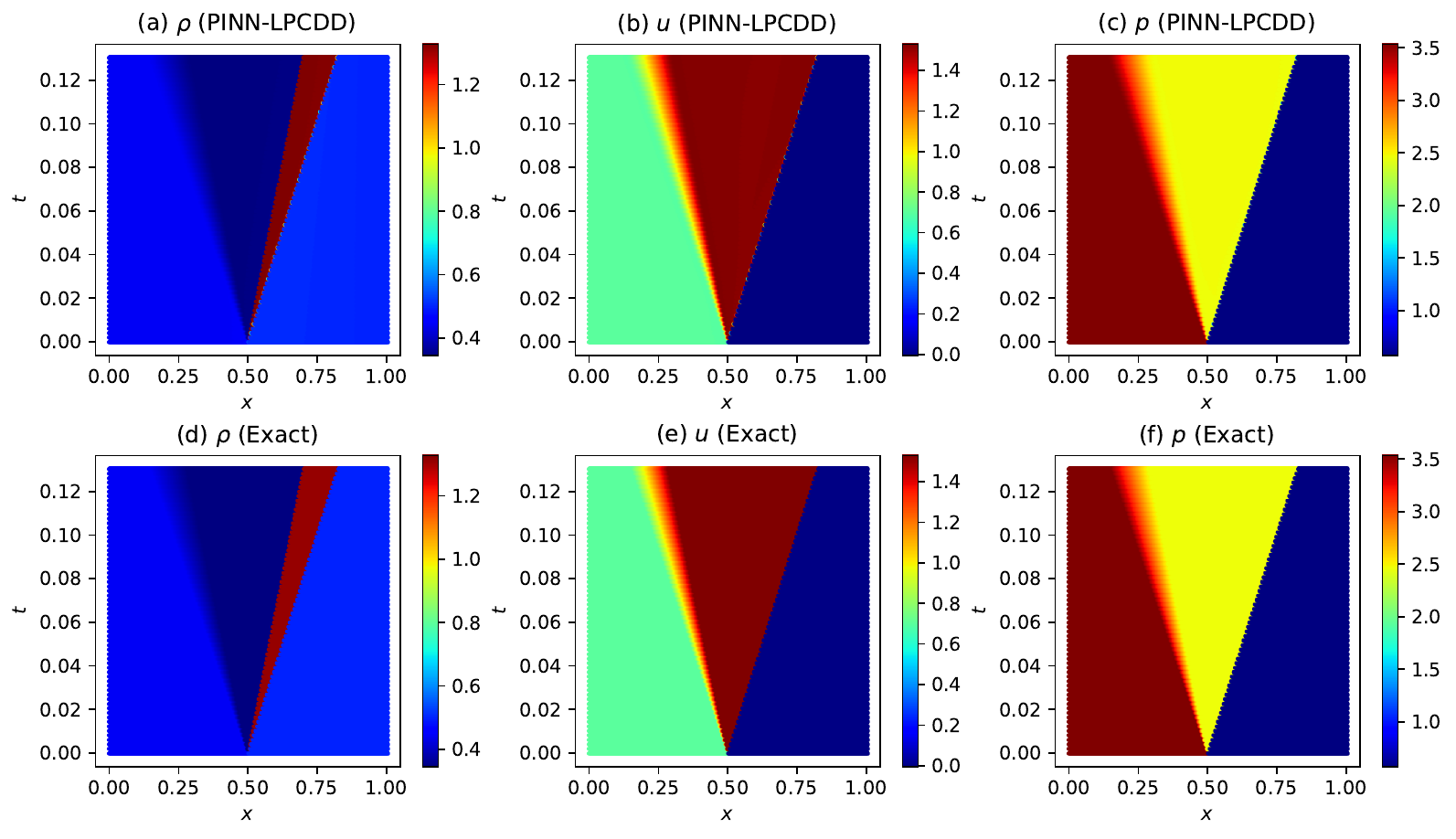}
    \includegraphics[width=\linewidth]{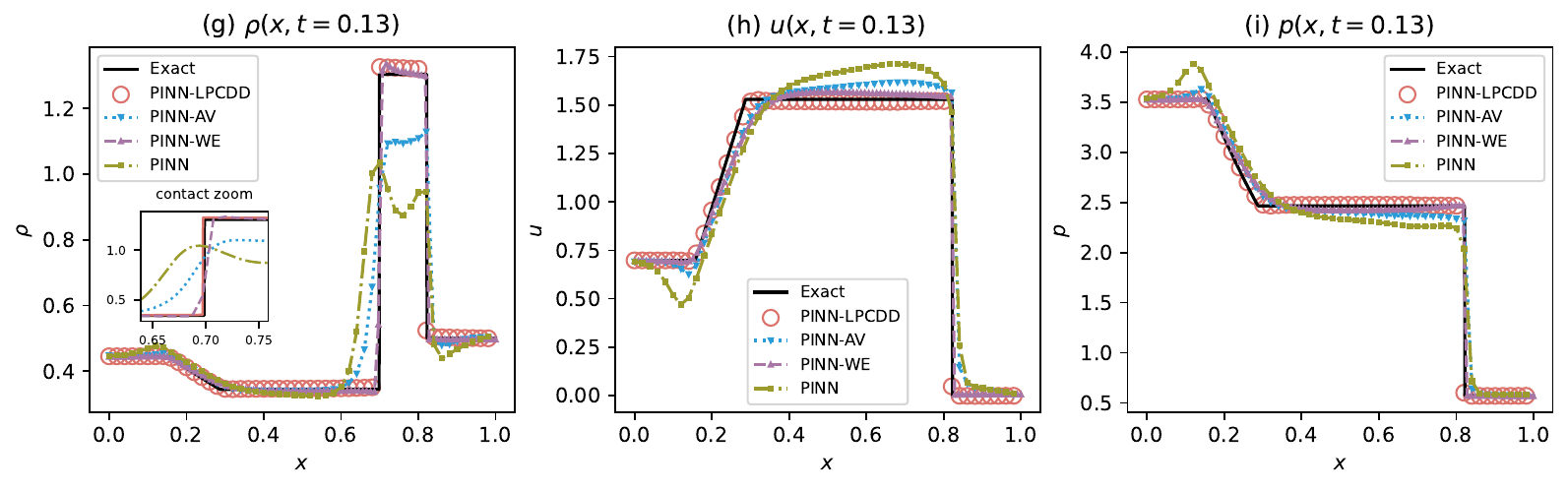}
    \caption{Results of 1D Lax problem (test case (1B)) with PINN-based methods.}
    \label{fig_compare_1B}
\end{figure}

From Fig.~\ref{fig_compare_1B}, it can be seen that in addition to the naturally sharp capture of the contact discontinuity, PINN-LPCDD also achieves clearly better approximation of the centered rarefaction wave than the other three models. Since the contact discontinuity in this test case is stronger than that in the Sod problem, the shortcomings of PINN-AV and PINN in approximating the contact discontinuity become more evident: they approximate the plateau to the right of the contact discontinuity rather poorly, leading to large errors there. Nevertheless, owing to the regularization of the artificial viscosity, PINN-AV still achieves better overall approximation than PINN. From the results in Table~\ref{tab_compare_1B}, conclusions similar to those of the Sod problem can be drawn: PINN-LPCDD achieves the smallest errors among all the models.

\begin{table}[H]
\caption{Comparison among PINN-based models in 1D Lax problem (test case (1B)).}\label{tab_compare_1B}
\small
\centering
\begin{tabular}{|c|cccc|c|}
\hline
Model & 
MSE ($\rho$) &
MSE ($u$) & 
MSE ($p$) &
MSE (avg) &
RE (avg) 
\\ \hline

PINN-LPCDD	& 4.875e-05	& 1.543e-04	& 1.769e-04	& 1.267e-04	& \bf{9.082e-03}	\\
PINN-AV	& 1.523e-02	& 2.791e-02	& 3.514e-02	& 2.609e-02	& 1.399e-01	\\
PINN-WE	& 4.375e-04	& 1.867e-03	& 2.351e-03	& 1.552e-03	& 2.998e-02	\\
PINN	& 3.742e-02	& 3.373e-02	& 4.085e-02	& 3.733e-02	& 1.858e-01	\\
\hline

\end{tabular}
\end{table}

\subsubsection{Double shock (DS) problem}

This test case examines the capability of PINN-based methods to approximate contact discontinuities and shock waves of varying strengths \cite{urban2025riemann}. The initial conditions are set to be

\begin{equation}
\mathrm{(1C)}\quad  
(\rho,u,p,\Gamma,B)=\begin{cases}
(1.0,\ 0.5,\ 1.0,\ 1.4,\ 0.0), & x<0.5; \\
(0.5,\ 0.0,\ 1.0,\ 1.4,\ 0.0), & x>0.5,
\end{cases}
\end{equation}
and we solve this problem on $(x,t)\in[0,1]\times[0,0.2]$.

For this test case, we also investigate the influence of the artificial viscosity coefficient $\nu$ on PINN-LPCDD. From this test case onward, all statistical results for 1D cases are computed on a grid with $N_x=1000$.

\begin{table}[H]
\caption{Comparison among PINN-based models in 1D double shock problem (test case (1C)).}\label{tab_compare_1C}
\small
\centering
\begin{tabular}{|c|c|cccc|c|}
\hline
Model & $\nu$ &
MSE ($\rho$) &
MSE ($u$) & 
MSE ($p$) &
MSE (avg) &
RE (avg) 
\\ \hline

\multirow{3}{*}{PINN-LPCDD}	& $2\times10^{-3}$	& 1.855e-04	& 4.994e-04	& 5.930e-04	& 4.260e-04	& 3.279e-02	\\
	& $2\times10^{-4}$	& 3.722e-05	& 1.024e-04	& 1.278e-04	& 8.914e-05	& \bf{1.491e-02}	\\
	& $2\times10^{-5}$	& 4.756e-04	& 3.962e-04	& 4.346e-04	& 4.355e-04	& 3.259e-02	\\
\hline
PINN-AV	& $2\times10^{-4}$	& 7.660e-04	& 1.441e-04	& 1.514e-04	& 3.539e-04	& 2.502e-02	\\
PINN	& $0$	& 2.580e-04	& 2.363e-04	& 2.777e-04	& 2.574e-04	& 2.504e-02	\\
\hline

\end{tabular}
\end{table}

\begin{figure}[H]
    \centering
    \includegraphics[width=\linewidth]{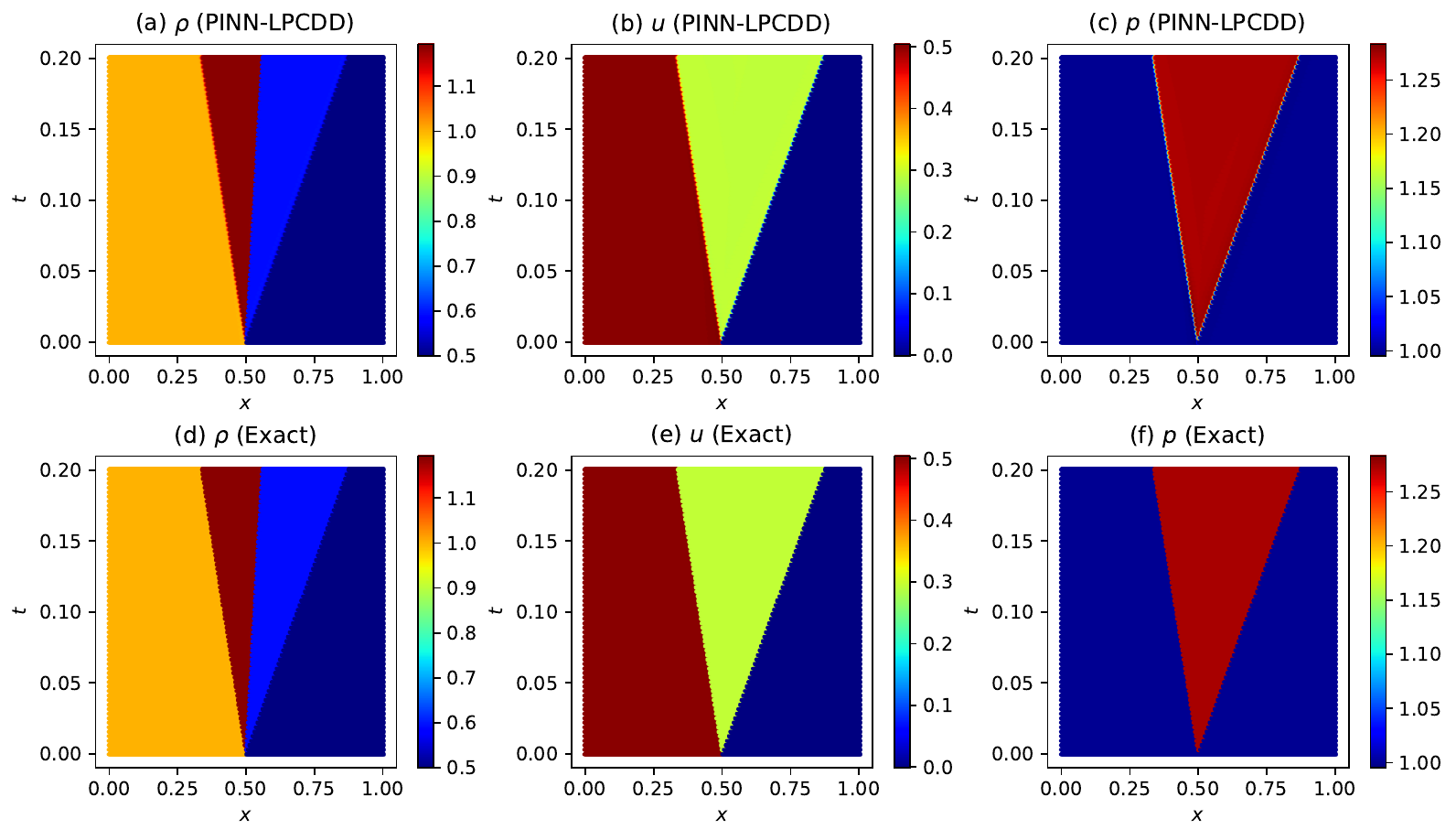}
    \includegraphics[width=\linewidth]{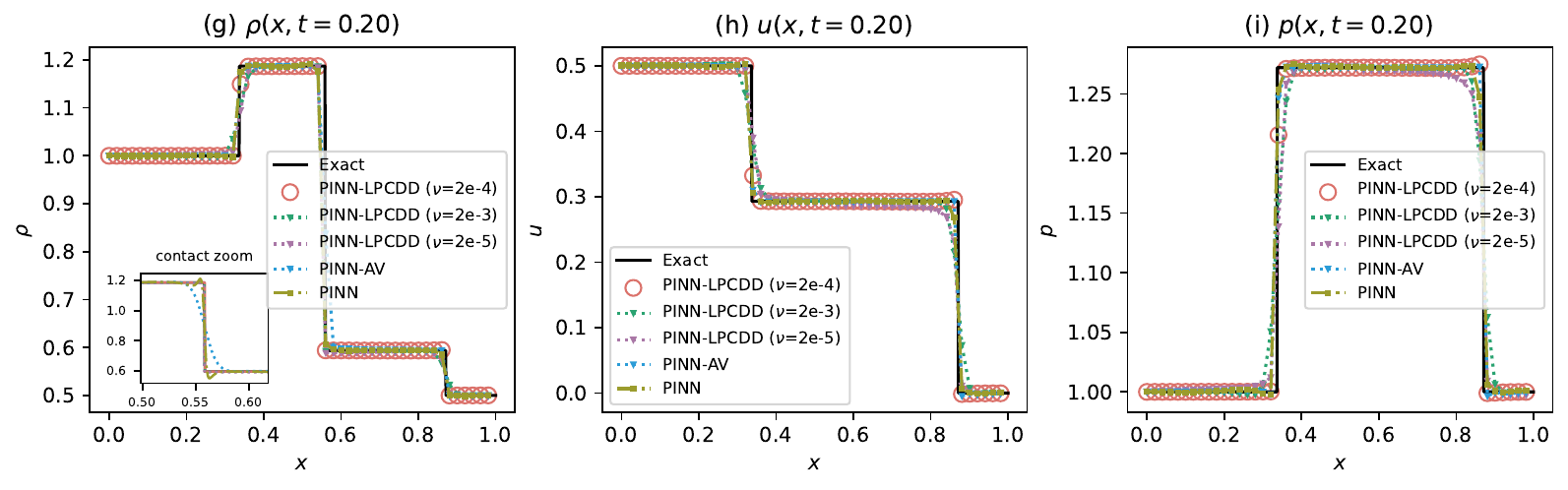}
    \caption{Results of 1D double shock problem (test case (1C)) with PINN-based methods.}
    \label{fig_compare_1C}
\end{figure}

It can be seen from Fig.~\ref{fig_compare_1C} and Table~\ref{tab_compare_1C} that when $\nu = 2 \times 10^{-3}$ is adopted, the viscosity is too large: the numerical solution is excessively smoothed and the errors of all variables increase noticeably. When $\nu = 2 \times 10^{-4}$, PINN-LPCDD achieves the best accuracy. When $\nu$ is further reduced to $2 \times 10^{-5}$, the effect of the artificial viscosity becomes negligible, indicating that such a small viscosity coefficient is no longer effective. Moreover, at the same viscosity coefficient ($\nu = 2 \times 10^{-4}$), PINN-LPCDD achieves better approximation than PINN-AV. The vanilla PINN also behaves rather competitively in this test case; however, its drawback lies in the numerical oscillations near the contact discontinuity, as shown in Fig.~\ref{fig_compare_1C}, whereas the methods with artificial viscosity do not suffer from this problem. 

\begin{figure}[H]
    \centering
    \includegraphics[width=\linewidth]{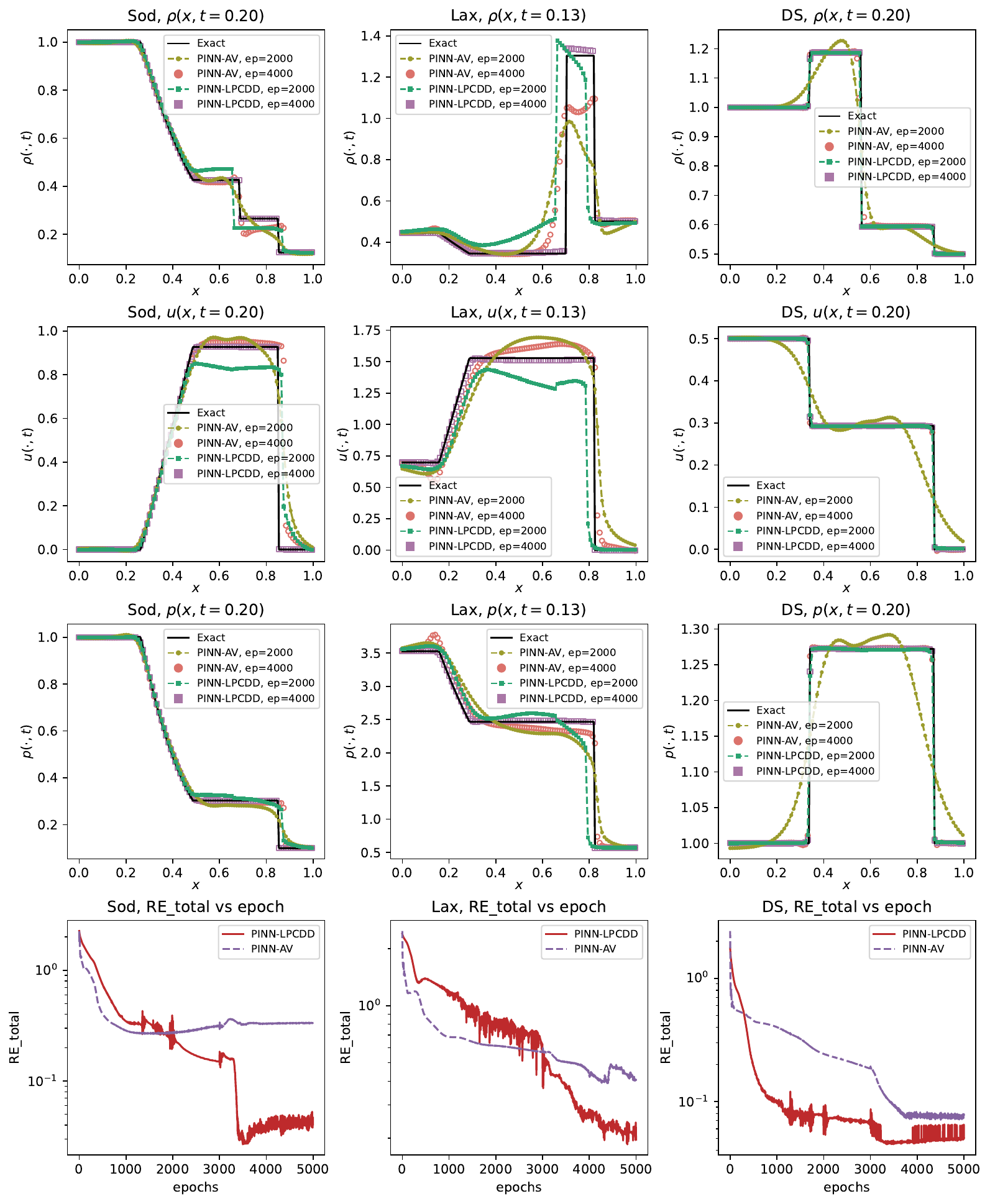}
    \caption{The performance history and the total RE history of PINN-LPCDD and PINN-AV in air-air problems.}
    \label{fig_compare_AA}
\end{figure}

At the end of this subsection, we conduct a more detailed comparison of the learning processes of PINN-AV and PINN-LPCDD. Fig.~\ref{fig_compare_AA} shows the performance of the two models at epochs 2000 and 4000, as well as the evolution curves of the total RE (i.e., the sum of the REs of the three primitive variables) with respect to the epoch. 

The following observations can be made: (1) Since PINN-LPCDD needs to learn the interface speed and couple the outputs of two neural networks at the interface, its approximation of the solution is significantly worse than that of PINN-AV in the early stage of training, mainly manifesting as a large discrepancy between the interface velocity and the true flow velocity. However, as training proceeds, the advantage of PINN-LPCDD in approximating contact discontinuities gradually emerges. (2) Numerical oscillations can be found in PINN-LPCDD during early training, but they can also be reduced as training progresses. 

\vspace{1cm}

\subsection{Air-helium Riemann problems}
\label{sec:air_helium}

In this subsection, we demonstrate the capability of PINN-LPCDD to solve problems involving two media with similar properties. In this case, the correct interface position must be identified to assign the corresponding states (i.e., $\Gamma$ and $B$) to each medium. The PINN-WE and PINN-AV approaches mentioned in the previous subsection cannot be directly applied to this problem. Moreover, the interface equation must be provided a priori if one uses cPINN/XPINN, which is impractical for real-world problems.

Therefore, we conduct only ablation studies in this subsection. We examine the necessity of the Fourier feature embedding, the RBF output layer, and the use of two separate sub-networks. We denote the variant that feeds the raw space-time coordinates directly into the fully-connected layers without the \textbf{\underline{F}}ourier feature \textbf{\underline{E}}mbedding as ``PINN-LPCDD w/o FE'', the variant that replaces the \textbf{\underline{R}}BF \textbf{\underline{L}}ayer with the standard linear readout as ``PINN-LPCDD w/o RL'', and the variant that does not use \textbf{\underline{S}}ub-\textbf{\underline{N}}etworks, i.e., uses only a single neural network, as ``PINN-LPCDD w/o SN''. Since removing the sub-network design or the RBF layer reduces the number of learnable parameters, for fairness we adjust the network width of the corresponding ablated models to 30, so that their learnable parameters are no fewer than those of PINN-LPCDD.

\subsubsection{Air with high pressure}
This test case examines the capability of PINN-LPCDD to approximate contact discontinuities and rarefaction waves of varying strengths \cite{xu2025unsupervised}. The initial conditions are set to be

\begin{equation}
\mathrm{(2A)}\quad  
(\rho,u,p,\Gamma,B)=\begin{cases}
(1.0,\ -1.0,\ 1.0,\ 1.4,\ 0.0), & x<0.6; \\
(1.0,\ 1.0,\ 0.1,\ 1.667,\ 0.0), & x>0.6,
\end{cases}
\end{equation}
and we solve this problem on $(x,t)\in[0,1]\times[0,0.2]$.

\begin{figure}[H]
    \centering
    \includegraphics[width=\linewidth]{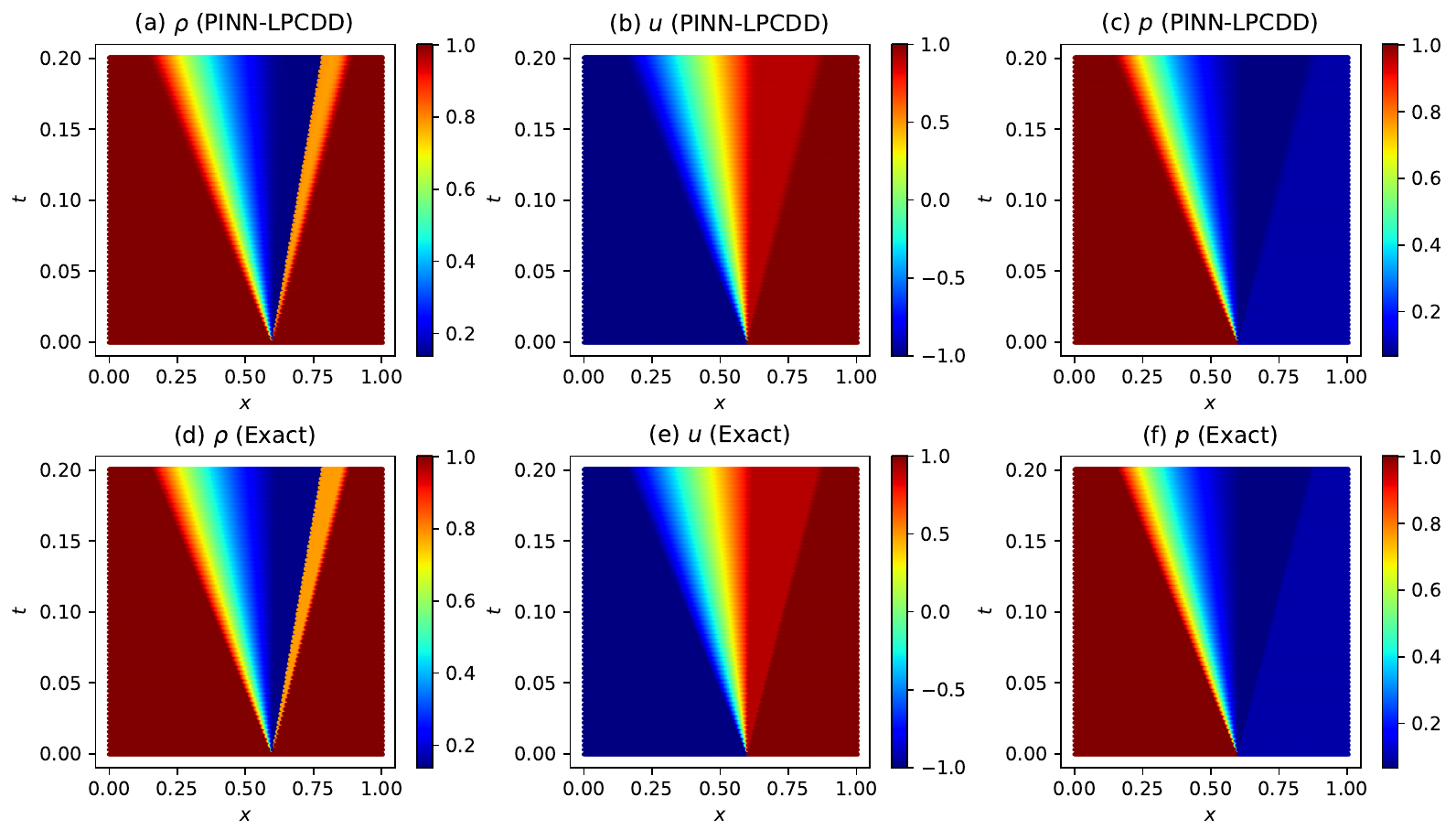}
    \includegraphics[width=\linewidth]{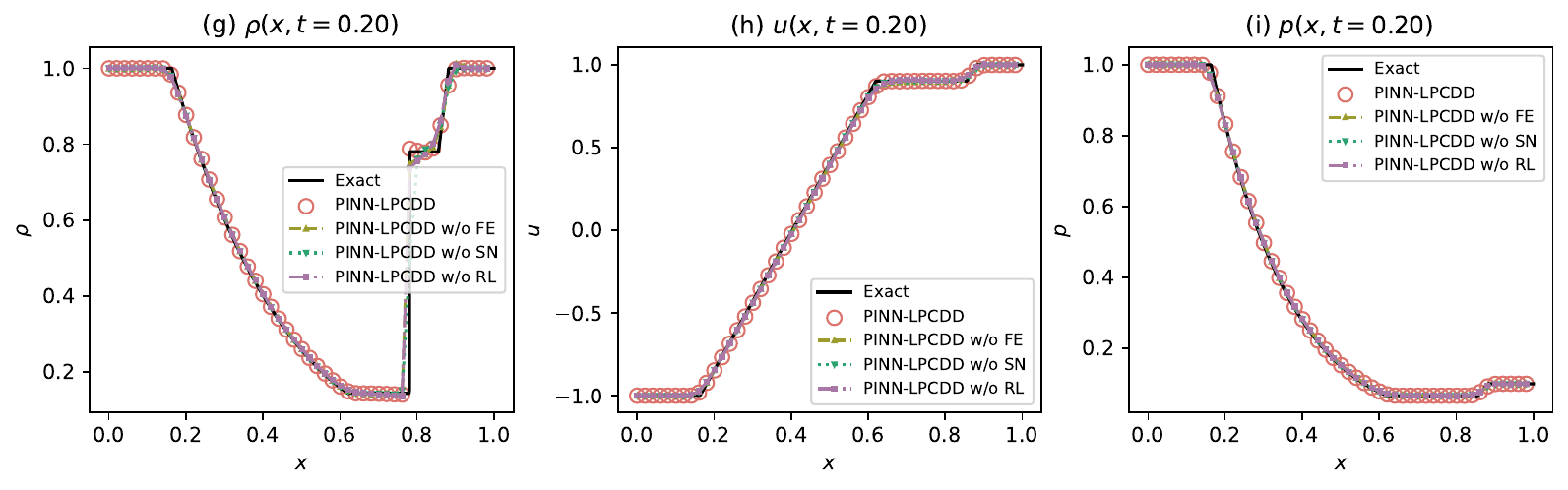}
    \caption{Results of 1D air-helium problem (test case (2A)) with PINN-LPCDD.}
    \label{fig_compare_2A}
\end{figure}

The difficulty of this test case lies in simultaneously distinguishing the contact discontinuity and the centered rarefaction wave within less than one-fifth of the spatial domain. As can be seen from Fig.~\ref{fig_compare_2A}, PINN-LPCDD achieves a better approximation of the step on the right side of the interface than the ablated models, which demonstrates the necessity of the Fourier feature embedding, the RBF layer, and the dual-network design. The result in Table~\ref{tab_compare_2A} further confirms this analysis.

\begin{table}[H]
\caption{Comparison among PINN-LPCDD variants in air-helium problem (test case (2A)).}\label{tab_compare_2A}
\small
\centering
\begin{tabular}{|c|cccc|c|}
\hline
Model & 
MSE ($\rho$) &
MSE ($u$) & 
MSE ($p$) &
MSE (avg) &
RE (avg) 
\\ \hline

PINN-LPCDD		& 4.665e-05	& 8.051e-05	& 1.936e-05	& 4.884e-05	& \bf{9.916e-03}	\\
PINN-LPCDD w/o FE	& 4.341e-04	& 1.111e-04	& 2.736e-05	& 1.909e-04	& 1.786e-02	\\
PINN-LPCDD w/o SN	& 9.474e-04	& 7.692e-05	& 2.309e-05	& 3.491e-04	& 2.163e-02	\\
PINN-LPCDD w/o RL	& 9.846e-05	& 1.289e-04	& 3.078e-05	& 8.605e-05	& 1.315e-02	\\
\hline

\end{tabular}
\end{table}

\subsubsection{Helium with high pressure}

The purpose of this test case is similar to that of the Lax problem, with the difference that two different materials must also be distinguished here \cite{xu2025unsupervised}. The initial conditions are set to be

\begin{equation}
\mathrm{(2B)}\quad  
(\rho,u,p,\Gamma,B)=\begin{cases}
(1.0,\ -1.0,\ 1.0,\ 1.4,\ 0.0), & x<0.5; \\
(1.0,\ 1.0,\ 10.0,\ 1.667,\ 0.0), & x>0.5,
\end{cases}
\end{equation}
and we solve this problem on $(x,t)\in[0,1]\times[0,0.08]$. 

\begin{figure}[H]
    \centering
    \includegraphics[width=\linewidth]{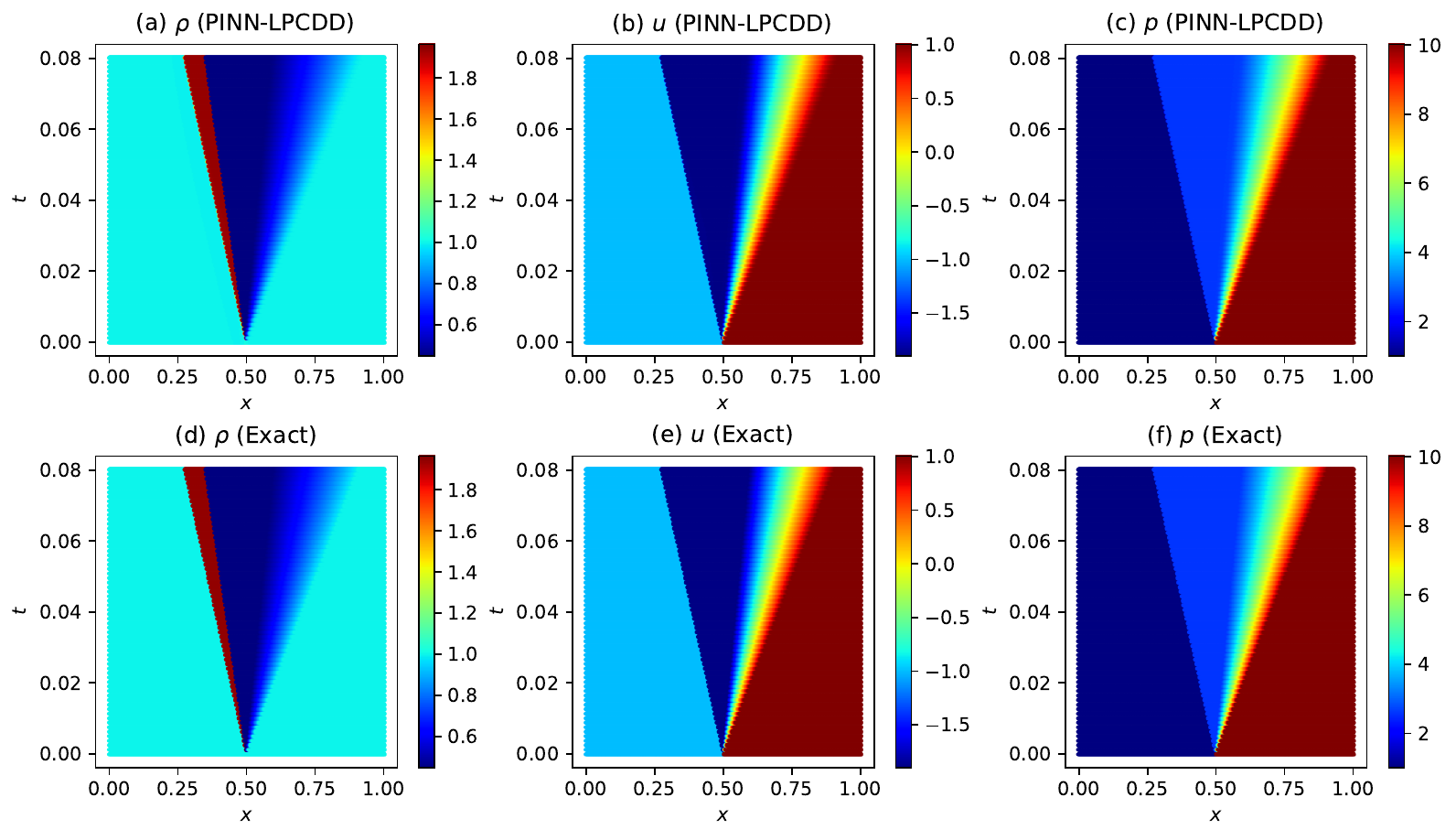}
    \includegraphics[width=\linewidth]{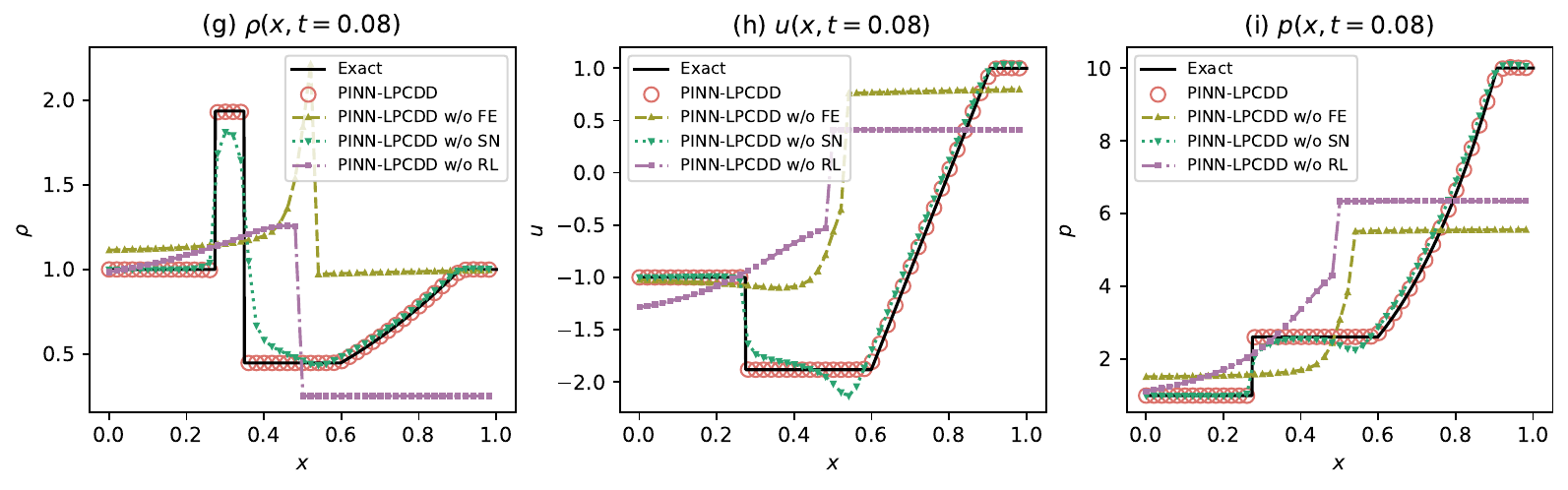}
    \caption{Results of 1D air-helium problem with helium high pressure (test case (2B)) with PINN-LPCDD.}
    \label{fig_compare_2B}
\end{figure}

The difficulty of this test case lies in approximating the plateau to the left of the interface. From Fig.~\ref{fig_compare_2B}, it can be seen that removing either the Fourier feature embedding or the RBF layer makes the model fail to correctly learn the interface position, resulting in poor solutions. When only the sub-network design is removed, the learned solution is excessively smeared near the interface, and the errors around the interface are relatively large. Only by incorporating all of these designs can PINN-LPCDD maximize its learning capability. The results in Table~\ref{tab_compare_2B} are consistent with this analysis.

\begin{table}[H]
\caption{Comparison among PINN-LPCDD variants in air-helium problem (test case (2B)).}\label{tab_compare_2B}
\small
\centering
\begin{tabular}{|c|cccc|c|}
\hline
Model & 
MSE ($\rho$) &
MSE ($u$) & 
MSE ($p$) &
MSE (avg) &
RE (avg) 
\\ \hline

PINN-LPCDD		& 2.160e-04	& 5.149e-04	& 2.132e-03	& 9.542e-04	& \bf{1.414e-02}	\\
PINN-LPCDD w/o FE	& 2.946e-01	& 1.396e+00	& 4.109e+00	& 1.933e+00	& 6.301e-01	\\
PINN-LPCDD w/o SN	& 1.674e-02	& 1.124e-02	& 4.738e-02	& 2.512e-02	& 8.807e-02	\\
PINN-LPCDD w/o RL	& 2.656e-01	& 1.318e+00	& 4.480e+00	& 2.021e+00	& 6.181e-01	\\
\hline

\end{tabular}
\end{table}

\subsection{Air-water Riemann problems}
\label{sec:air_water}

In this subsection, we demonstrate the capability of PINN-LPCDD to solve problems involving two media with significant differences, which poses difficulties for neural network approximation. Under such circumstances, the scaling transformation mentioned in Subsection~\ref{sec:preprocessing} becomes crucial, and this will be discussed in detail below.

\subsubsection{1D plane flow}

The initial conditions for this test case \cite{xu2025unsupervised} are set to be

\begin{equation}
\mathrm{(3A)}\quad  
(\rho,u,p,\Gamma,B)=\begin{cases}
(1.0,\ 0.0,\ 10000.0,\ 1.4,\ 0.0), & x<0.4; \\
(1.0,\ 0.0,\ 1.0,\ 4.4,\ 6000.0), & x>0.4,
\end{cases}
\label{ic_3A}
\end{equation}
and we solve this problem on $(x,t)\in[0,1]\times[0,0.002]$. 

\begin{figure}[H]
    \centering
    \includegraphics[width=\linewidth]{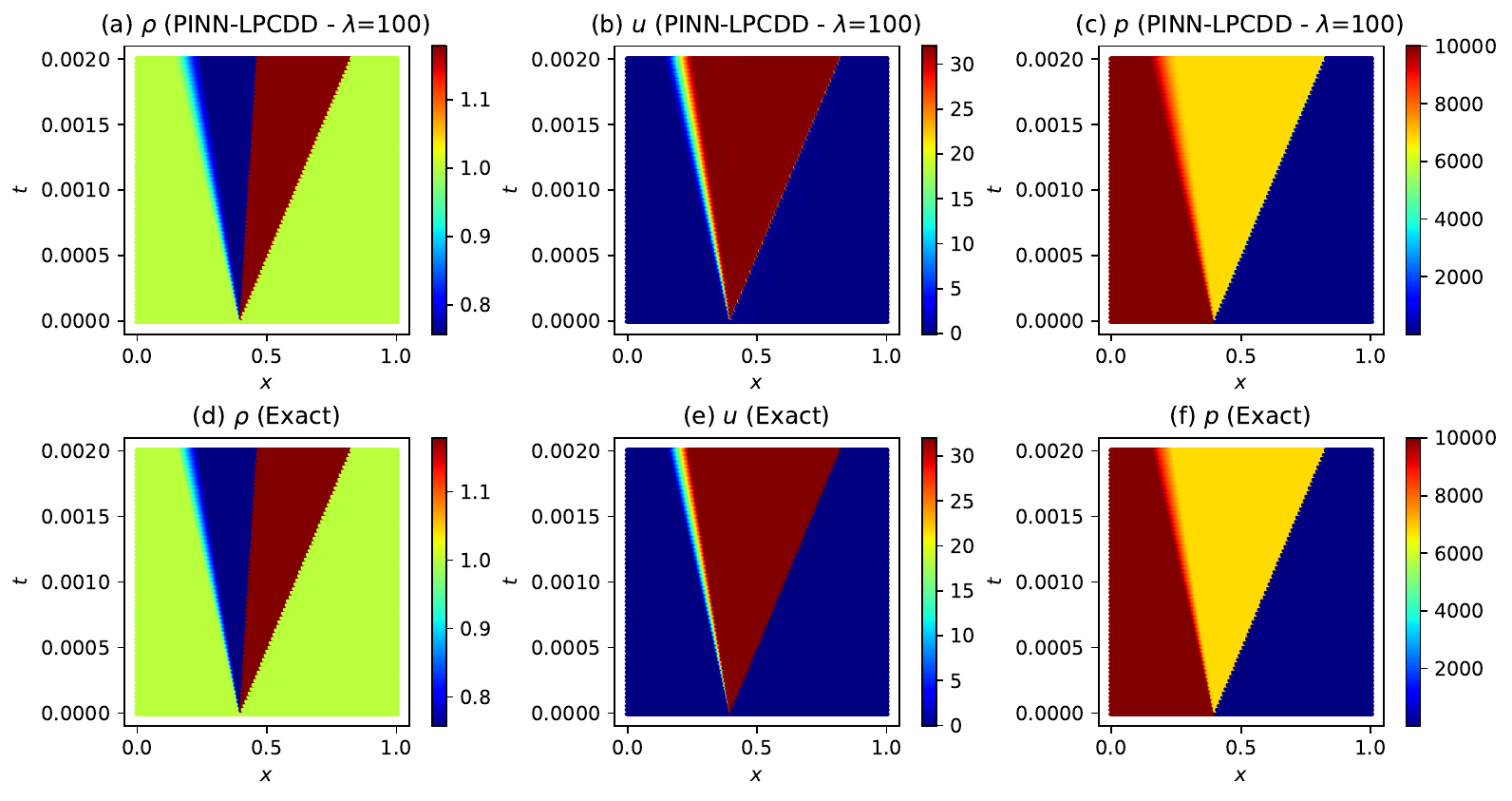}
    \includegraphics[width=\linewidth]{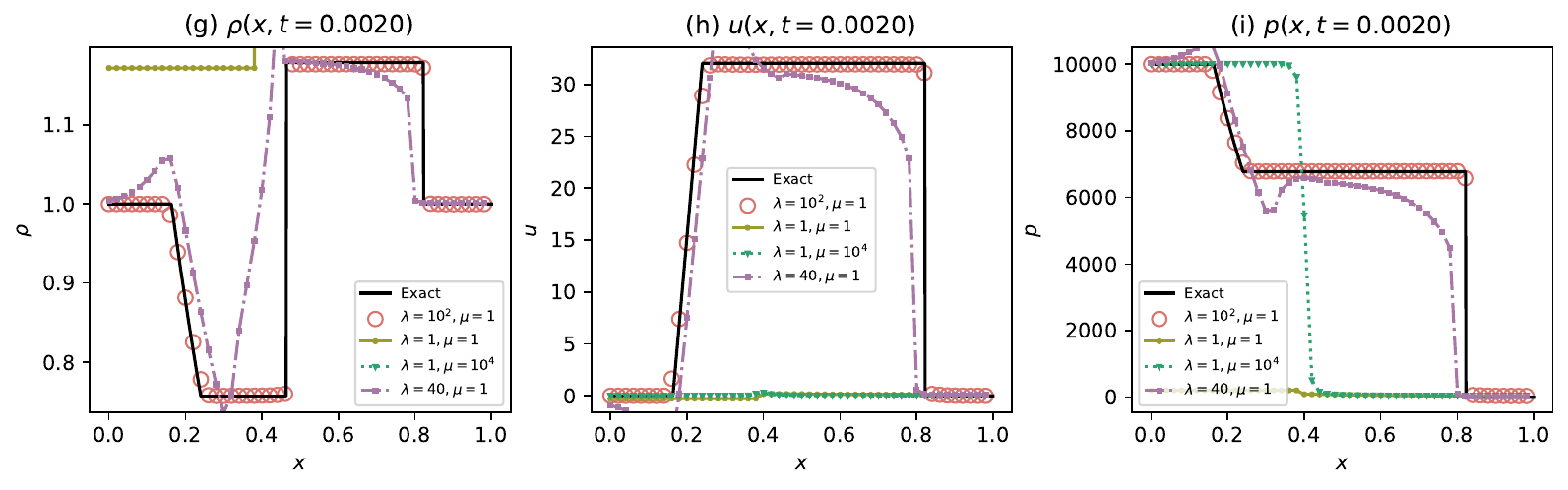}
    \caption{Results of 1D air-water plane flow problem (test case (3A)) with PINN-LPCDD.}
    \label{fig_compare_3A}
\end{figure}

We set up four different strategies for selecting $(\lambda, \mu)$, and the results are shown in Fig.~\ref{fig_compare_3A}. It is found that PINN-LPCDD performs best when $\lambda = 10^2$. Combined with Eq.~\eqref{scaled_fields}, it can be seen that under this transformation, the true $\rho$, $u$, and $p$ all fall within the interval $[0, 1]$, and the gradients of the physical quantities are also reduced, making it easier for the neural network to approximate the solution. When $\lambda = 40$ is selected, the shape of the predicted solution curve resembles the exact solution, but exhibits relatively large dissipation. When $\mu = 10^4$ is selected, the solution cannot be effectively learned, similar to the case without scaling. This may be because, although the initial condition can be transformed to the $[0, 1]$ interval, it is still an order of magnitude away from the maximum velocity, leading to ill-conditioning in the learning process. 

The results in Table~\ref{tab_compare_3A} show that selecting $\lambda = 10^2$ yields good performance, outperforming the model with $\lambda = 40$, while the models with $\lambda = 1$, i.e., without proper velocity scaling, result in very large errors.

\begin{table}[H]
\caption{Comparison among different scaling strategies for PINN-LPCDD in 1D air-water plane flow problem (test case (3A)).}\label{tab_compare_3A}
\small
\centering
\begin{tabular}{|c|cccc|c|}
\hline
$(\lambda, \mu)$ & 
MSE ($\rho$) &
MSE ($u$) & 
MSE ($p$) &
MSE (avg) &
RE (avg) 
\\ \hline
$(10^2,1)$	& 2.032e-05	& 4.382e-01	& 1.046e+04	& 3.488e+03	& \bf{1.522e-02}	\\
$(40,1)$	& 1.635e-02	& 5.383e+01	& 2.022e+06	& 6.740e+05	& 2.082e-01	\\
$(1,10^4)$	& 1.060e+05	& 6.217e+02	& 2.025e+07	& 6.785e+06	& 1.077e+02	\\
$(1,1)$	& 1.232e+01	& 6.227e+02	& 4.663e+07	& 1.554e+07	& 1.815e+00	\\
\hline

\end{tabular}
\end{table}
\vspace{1cm}

\subsubsection{2D cylindrically symmetric flow (explosion under water)}

This test case examines the capability of our method to generalize to 2D problems. The initial conditions are set to be

\begin{equation}
\mathrm{(3B/3C)}\quad
(\rho,u,v,p,\Gamma,B)=\begin{cases}
(1.0,\ 0.0,\ 0.0, \ 10000.0,\ 1.4,\ 0.0), & x^2+y^2<0.4^2; \\
(1.0,\ 0.0,\ 0.0,\ 1.0,\ 4.4,\ 6000.0), & x^2+y^2>0.4^2.
\end{cases}
\end{equation}

We adopt two approaches to solve this problem: 
\begin{itemize}
\item \emph{Test case (3B)}: transform the problem into a 1D problem with a source term, i.e., solving Eq.~\eqref{euler_pde} with $\alpha=1$, where the initial condition is consistent with Eq.~\eqref{ic_3A}. The computational domain is set to $(x,t) \in [0,1] \times [0,0.002]$; 
\item \emph{Test case (3C)}: solve the 2D problem directly. Since this case is symmetric about the $x$-axis, we compute only the upper half-plane and set the computational domain to $(x,y,t)\in[-1,1]^2\times[0,0.002]$, apply a reflecting boundary condition on the bottom surface, and impose Dirichlet boundary conditions on the remaining surfaces.
\end{itemize}

In the preceding test cases, the interface velocity is constant and hence the interface can be represented by the single parameter $u_*$. From this subsection onward, however, the interface velocity varies with time, so the interface needs to be sampled more densely. For test case (3B), the interface is parameterized by the interface velocities $\{u_{*,k}\}_{k=1}^{N_t}$ at all time levels. For test case (3C), we sample $\{(x_{0,j}, y_{0,j})\}=\{(0.4\cos(\frac{j\pi}{J}), 0.4\sin(\frac{j\pi}{J}))\}_{j=0}^{J}$ with $J=17$ from the initial interface and track the velocities $\{u_{*,k,j}, v_{*,k,j}\}_{k=1,j=0}^{k=N_t,j=J}$ in the neural network as mentioned in Subsection~\ref{sec:rls}. Additionally, here we set $(\lambda,\mu)=(10^2, 1)$, the best parameters found in test case (3A), and the artificial viscosity coefficient is set to $\nu=5\times10^{-4}$ in this subsection.

\begin{figure}[H]
    \centering
    \includegraphics[width=\linewidth]{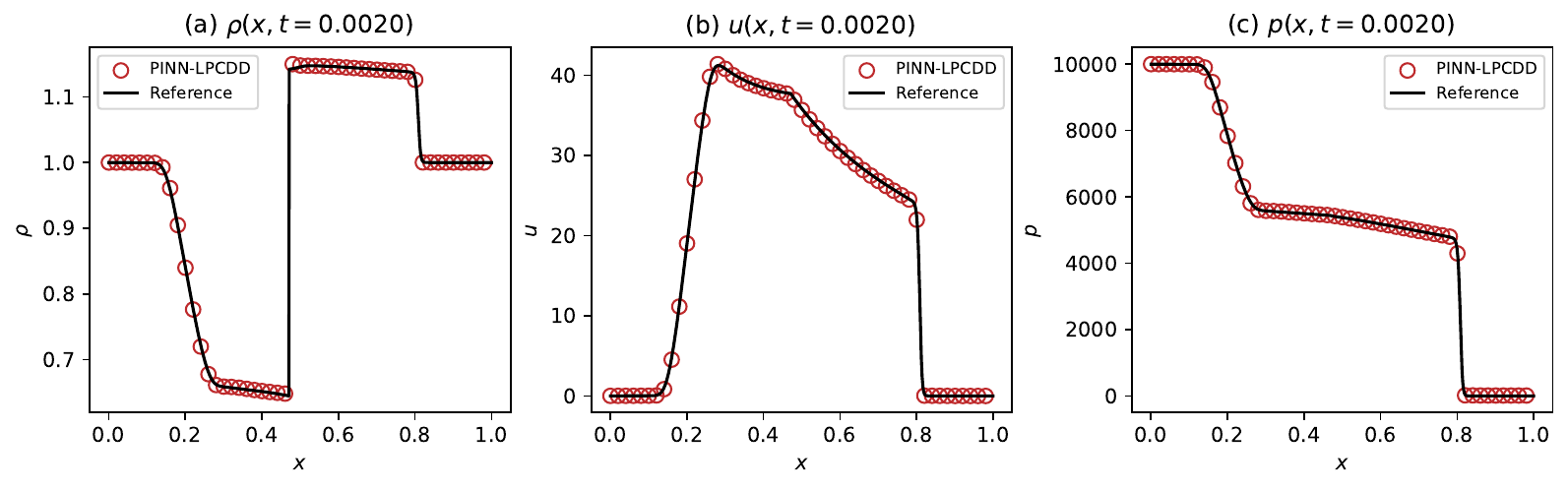}
    \caption{Results of 2D cylindrically symmetric explosion problem (test case (3B)) with PINN-LPCDD.}
    \label{fig_compare_3B}
\end{figure}

\begin{figure}[H]
    \centering
    \includegraphics[width=\linewidth]{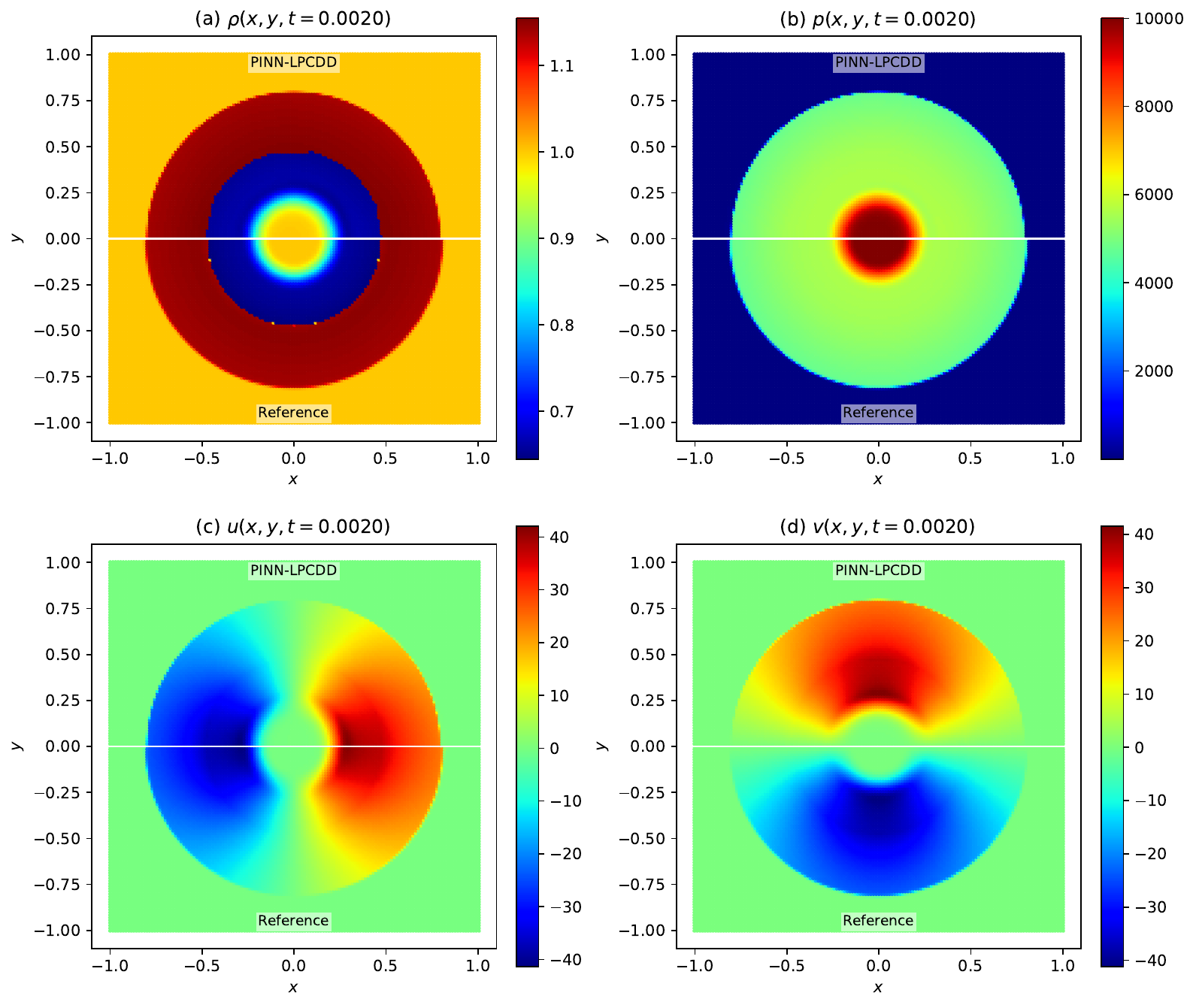}
    \caption{Results of 2D cylindrically symmetric problem (test case (3C)) with PINN-LPCDD.}
    \label{fig_compare_3C}
\end{figure}

\begin{figure}[H]
    \centering
    \includegraphics[width=\linewidth]{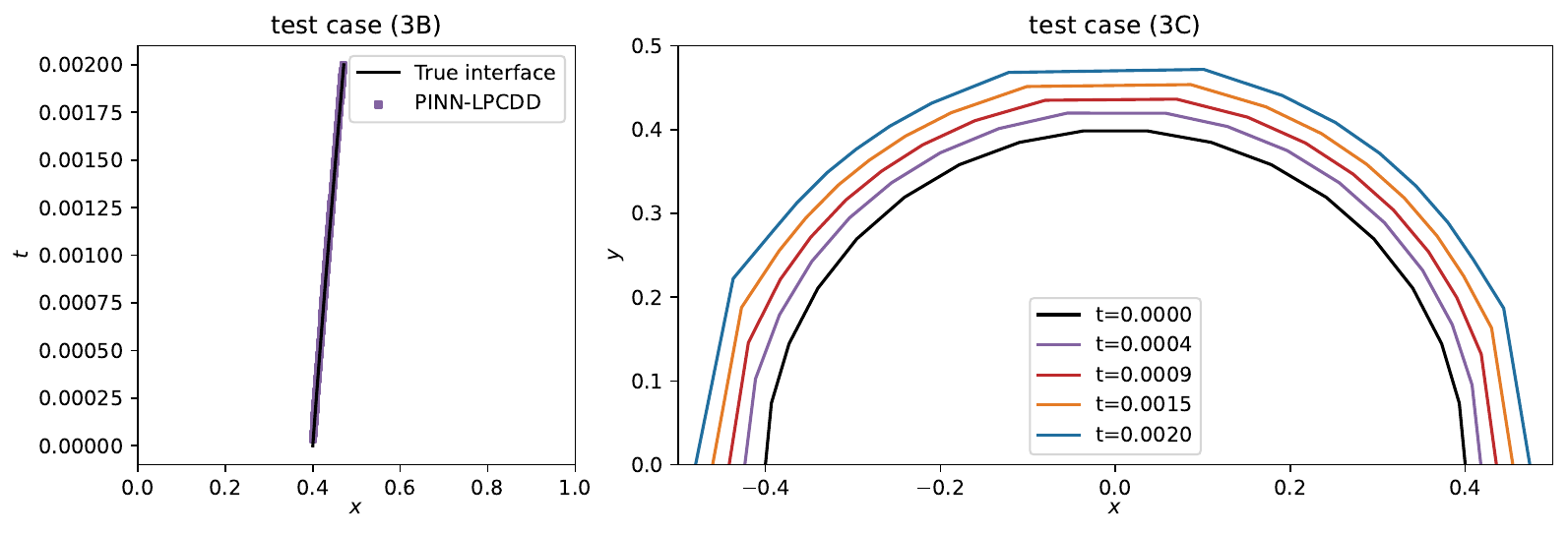}
    \caption{Interface track of 2D cylindrically symmetric explosion problem.}
    \label{fig_compare_3BC_interface_track}
\end{figure}

Results in Figs.~\ref{fig_compare_3B}, \ref{fig_compare_3C}, and Table~\ref{tab_compare_3BC} show that both approaches are capable of solving the underwater explosion problem.  Since the grid used for the 2D problem is coarser than that for the 1D problem, its average RE is larger than that of the 1D case. Additionally, Fig.~\ref{fig_compare_3BC_interface_track} shows the interface predicted by the neural network in the underwater explosion test case. It is computed according to Eqs.~\eqref{interface1D} and \eqref{interface2D}, using the learned parameter values of the neural network. It can be observed that when the 1D formulation is used, the interface is well predicted; however, when the 2D formulation is used, the interface no longer maintains a circular shape well, although this does not significantly affect the final results, as seen in Fig.~\ref{fig_compare_3C}.

\begin{table}[H]
\caption{Errors of PINN-LPCDD in 2D air-water explosion problems (test cases (3B) and (3C)).}\label{tab_compare_3BC}
\small
\centering
\begin{tabular}{|c|cccc|c|}
\hline
Test Case & 
MSE ($\rho$) &
MSE ($u$) & 
MSE ($p$) &
MSE (avg) &
RE (avg) 
\\ \hline

(3B) 	& 2.849e-04	& 7.791e-01	& 2.640e+04	& 8.800e+03	& 2.593e-02	\\
(3C)   	& 9.739e-04	& 7.162e+00	& 2.461e+05	& 8.203e+04	& 7.196e-02	\\
\hline

\end{tabular}
\end{table}

\subsection{Shock-interface interaction}
\label{sec:shock_interface}

In this subsection, we test PINN-LPCDD for solving multiphysics processes and distinguishing among the various wave systems produced by shock-interface interactions.

\begin{figure}[H]
    \centering
    \includegraphics[width=\linewidth]{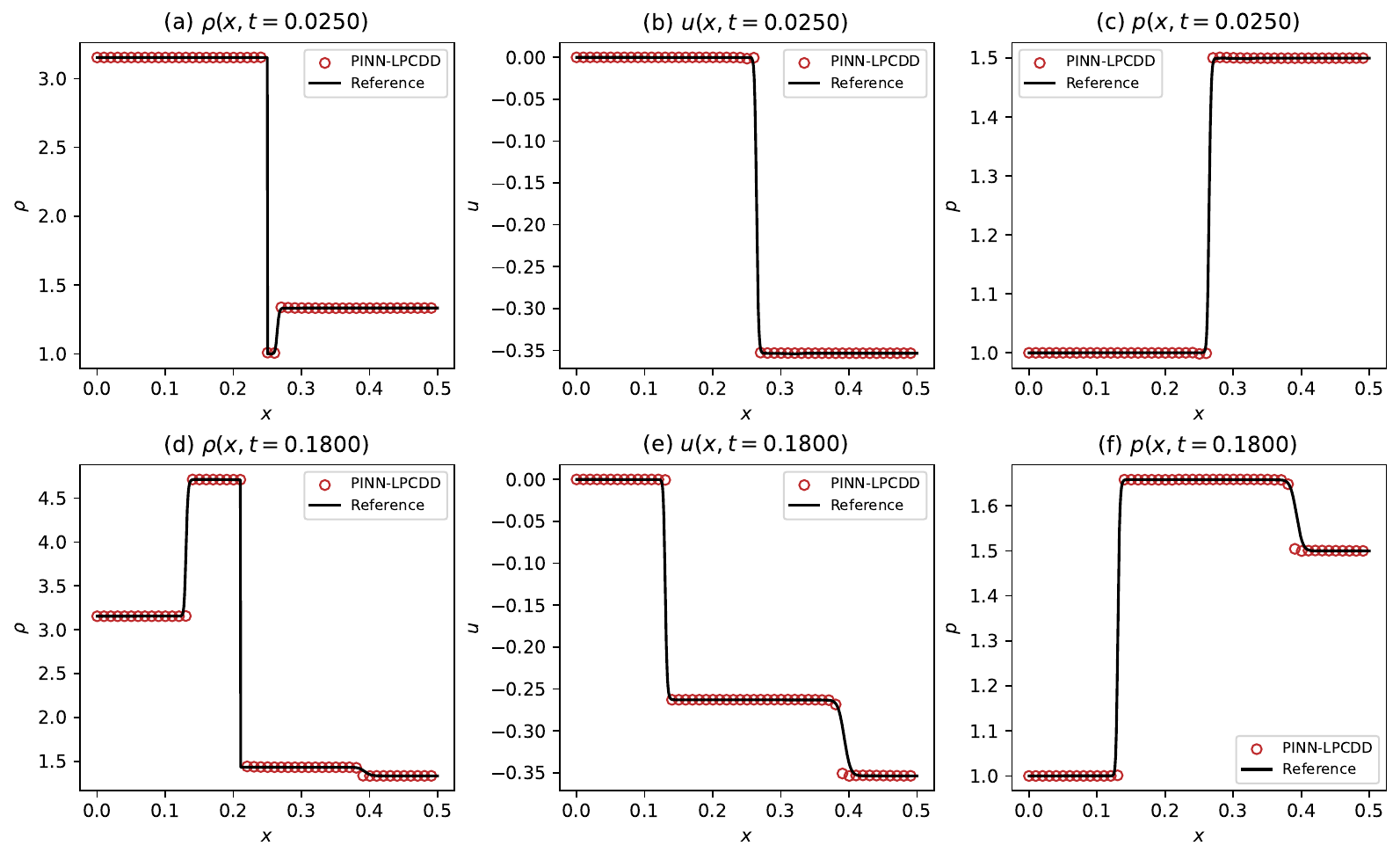}
    \caption{Results of 1D shock-interface interaction problem (test case (4A)) with PINN-LPCDD.}
    \label{fig_compare_4A}
\end{figure}

We test a 1D case first: the interaction between a left-traveling shock with Mach number $Ma=1.195$ in air and the air--R22 interface. The initial condition is set to be

\begin{equation} \mathrm{(4A)}\quad
(\rho,u,p,\Gamma,B)=\begin{cases}
(3.1538,\ 0.0,\ 1.0,\ 1.249,\ 0.0), & x<0.25;\\
(1.3333,\ -0.3535,\ 1.5,\ 1.4,\ 0.0), & x>0.30;\\
(1.0,\ 0.0,\ 1.0,\ 1.4,\ 0.0), & \mathrm{otherwise}
,
\end{cases}
\end{equation}
and we solve this problem on $(x,t)\in[0,0.5]\times[0,0.18]$. Additionally, we set $(\lambda,\mu)=(3,0.5)$, and first train with Adam for 1200 epochs and then with SOAP for 8800 additional epochs in this subsection. The interface parameterization is the same as that in test case (3B).

This test case involves two physical processes: (1) for $t < \frac{\sqrt{2}}{40}$, the initial shock wave moves toward the interface at a constant speed with $Ma = 1.195$; (2) for $t \geq \frac{\sqrt{2}}{40}$, the shock wave impacts the interface, generating a transmitted strong shock wave inside the R22 region and a reflected weak shock wave in the air. The results shown in Fig.~\ref{fig_compare_4A} indicate that PINN-LPCDD yields accurate solutions in both phases, prior to and following the shock impact.

We next extend the test case to the 2D setting \cite{haas1987interaction}: a cylindrical R22 bubble located at the center of the domain, which would be collapsed by a left-traveling shock with Mach number $Ma=1.195$. The initial condition is set to be

\begin{equation} \mathrm{(4B)}\quad
(\rho,u,p,\Gamma,B)=\begin{cases}
(3.1538,\ 0.0,\ 1.0,\ 1.249,\ 0.0), & x^2+y^2<0.25^2;\\
(1.3333,\ -0.3535,\ 1.5,\ 1.4,\ 0.0), & x>0.30;\\
(1.0,\ 0.0,\ 1.0,\ 1.4,\ 0.0). & \mathrm{otherwise}
.
\end{cases}
\end{equation}
Since this case is symmetric about the $x$-axis, we compute only the upper half-plane as in test case (3C). Moreover, because the initial shock has not crossed the central axis of the bubble by the terminal time, we compute only the right half of the bubble. Finally, we set the computational domain to $(x,y,t)\in[0,0.5]^2\times[0,0.18]$, and apply the outflow condition on the top boundary, the symmetry condition on the bottom boundary, and Dirichlet conditions on the left and right boundaries. We sample $\{(x_{0,j}, y_{0,j})\}=\{(0.25\cos(\frac{j\pi}{2J}), 0.25\sin(\frac{j\pi}{2J}))\}_{j=0}^{J}$, $J=17$, from the initial interface for parameterization.

\begin{figure}[H]
    \centering
    \includegraphics[width=0.85\linewidth]{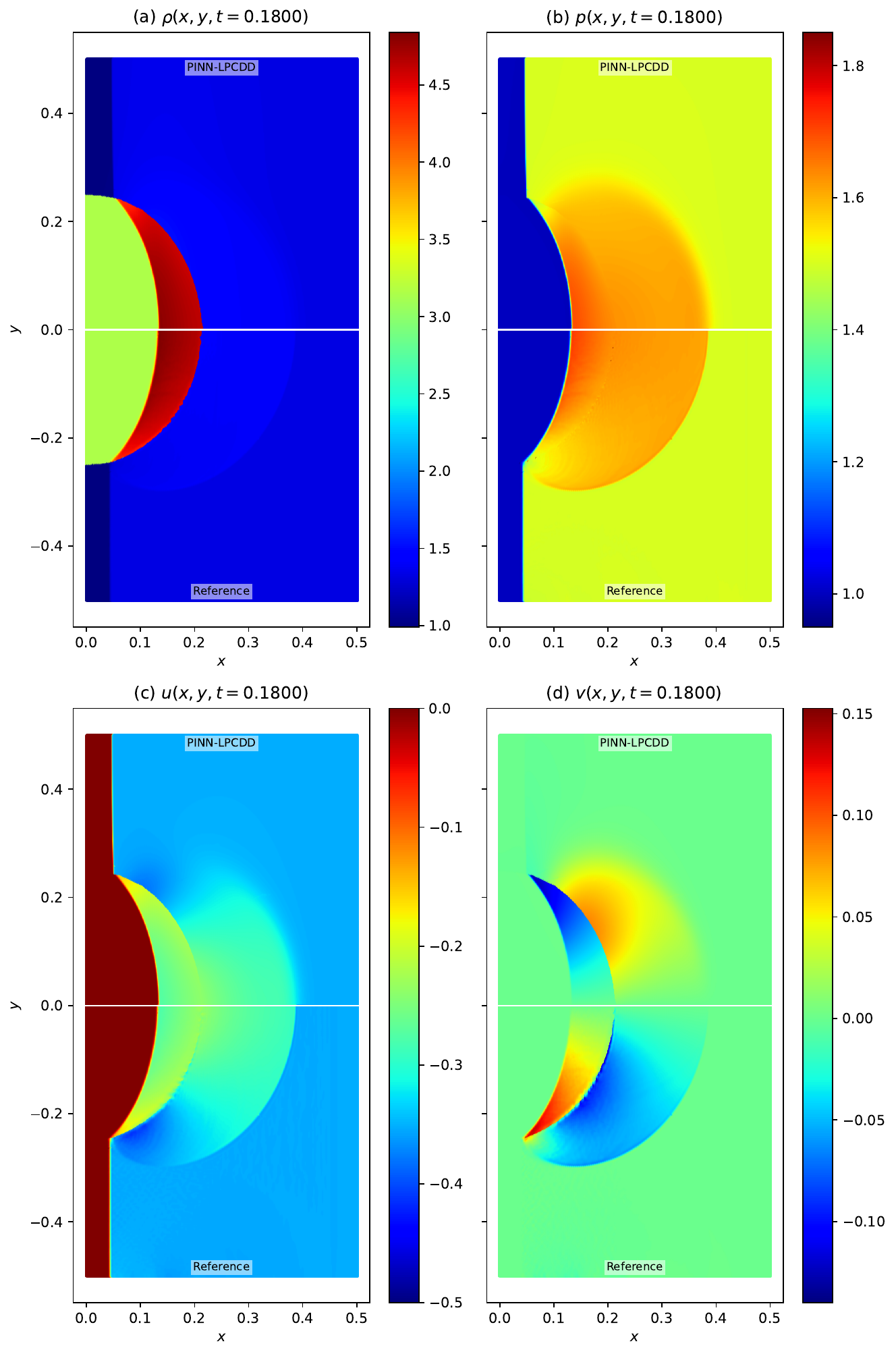}
    \caption{Results of 2D shock-interface interaction problem (test case (4B)) with PINN-LPCDD.}
    \label{fig_compare_4B}
\end{figure}

As can be seen from Fig.~\ref{fig_compare_4B}, PINN-LPCDD captures the left-traveling shock in both the air and R22 regions, although there are slight discrepancies in the shock positions. Furthermore, the reflected shock in the air region is smeared to a certain extent. The results in Table~\ref{tab_compare_4AB} further confirm that PINN-LPCDD performs well on the 1D shock–bubble interaction problem, while its performance on the 2D problem is less satisfactory. Note that the relatively large average RE is mainly attributed to the large RE of $v$: since the values of $v$ are close to zero in most of the domain, even small absolute errors lead to large relative errors.

\begin{table}[H]
\caption{Errors of PINN-LPCDD in shock-interface interaction problems (test cases (4A) and (4B)).}\label{tab_compare_4AB}
\small
\centering
\begin{tabular}{|cc|cccc|c|}
\hline
Test Case & Indicator & $\rho$ & $u$ & $v$ & $p$ & avg \\
\hline
\multirow{2}{*}{(4A)} & MSE & 9.607e-03	& 3.475e-04	& ---	& 1.858e-03	& 3.937e-03 \\
 & RE & 3.640e-02 & 7.440e-02 & --- & 2.920e-02 & 4.660e-02 \\
\hline
\multirow{2}{*}{(4B)} & MSE & 2.676e-02	& 6.346e-04	& 9.677e-05	& 2.039e-03	& 7.384e-03 \\
 & RE & 8.310e-02 & 8.160e-02 & 3.808e-01 & 3.080e-02 & 1.441e-01 \\
\hline
\end{tabular}
\end{table}

At the end of this section, we summarize the interface velocity learned by PINN-LPCDD for the 1D test cases in Table~\ref{tab_summary}. It is found that PINN-LPCDD learns the interface velocity well in the 1D test cases, with the relative $L^2$ error with respect to the true interface velocity ranging from $10^{-4}$ to $10^{-2}$.

\begin{table}[H]
\caption{Predictions of the interface velocity at the final time for the 1D test cases.}\label{tab_summary}
\small
\centering
\begin{tabular}{|c|cccc|}
\hline
Test Case & (1A) & (1B) & (1C) & (2A) \\
\hline
Exact & 9.275e-01 & 1.529e+00 & 2.929e-01 & 9.027e-01 \\
Prediction & 9.276e-01 & 1.515e+00 & 2.931e-01 & 9.037e-01 \\
RE & 1.614e-04 & 8.695e-03 & 5.980e-04 & 1.088e-03 \\
\hline
Test Case & (2B) & (3A) & (3B) & (4A) \\
\hline
Exact & $-1.883$e+00 & 3.203e+01 & 3.772e+01 & $-2.629$e-01 \\
Prediction & $-1.880$e+00 & 3.189e+01 & 3.765e+01 & $-2.628$e-01 \\
RE & 1.652e-04 & 4.616e-03 & 1.765e-03 & 3.874e-04 \\
\hline
\end{tabular}
\end{table}

In addition, the training time, training speed, and GPU memory usage of each test case are summarized in Table~\ref{tab_runtime} in Appendix~\ref{app:runtime}, where the training cost of PINN-LPCDD is discussed in detail. In particular, it is found that the computational (training) time of PINN-LPCDD on the 2D test case (4B) is already close to that of traditional methods.

\section*{Conclusion}
We proposed PINN-LPCDD, a physics-informed neural network framework tailored for multi-medium compressible flow problems, to address the difficulties of neural networks in resolving discontinuities and tracking moving medium interfaces. To overcome the challenge of approximating discontinuous solutions, we decompose the computational domain into two sub-domains separated by the medium interface and deploy two sub-networks, each responsible for one medium, so that the contact discontinuity is represented explicitly rather than smeared. To locate the interface without any a priori information, we introduce a learnable interface parameterization, which treats the interface velocities as learnable parameters and couples the two sub-networks through the interface continuity loss. To alleviate the ill-conditioning caused by large pressure ratios, we adopt a data preprocessing strategy, i.e., the two-parameter scaling transformation of the primitive variables. The research encompassed the following aspects:

\begin{itemize}
    \item The design of the PINN-LPCDD solver. To resolve discontinuities, we adopted the subdomain-wise dual-network architecture for contact discontinuities, together with a shock representation mechanism that combines the Fourier feature embedding and the RBF output layer; to track the interface, we devised the learnable interface parameterization, which learns the interface moving speed via the improved interface loss; to handle data scale disparities, we introduced the scaling transformation of the primitive variables, together with the corresponding transformed equations.
    \item The numerical validation on a series of multi-medium Riemann problems. We conducted extensive experiments covering single-medium problems (Sod, Lax, double shock), two-medium problems with similar properties (air-helium and air-R22), and two-medium problems with extreme pressure ratios (air-water), together with the extension to two-dimensional problems.
\end{itemize}

According to the results of the numerical experiments, we draw the following conclusions:

\begin{itemize}
    \item The subdomain-wise dual-network strategy of PINN-LPCDD avoids the smearing of the contact discontinuity, since each medium is approximated by a dedicated sub-network without being smoothed across the interface. Compared with the artificial viscosity based approach (PINN-AV), PINN-LPCDD achieves markedly smaller errors at the contact discontinuity, and the numerical oscillations, if any, gradually diminish during training.
    \item The learnable interface parameterization enables PINN-LPCDD to learn the interface moving speed automatically, without requiring the interface equation a priori as in XPINN. The learned interface velocity agrees well with the exact value.
    \item The data preprocessing strategy is indispensable for problems with large pressure ratios. The ablation studies confirm that removing either the Fourier feature embedding or the dual sub-network setting degrades the accuracy, and that an appropriate scaling parameter (e.g., $\lambda=100$ in the air-water problems) is crucial for effective learning.
\end{itemize}

The current limitations of this work and the future research directions are as follows:

\begin{itemize}
    \item The interface parameterization represents the interface by discrete marker points advanced with learnable velocities, which has difficulty in handling interfaces with complex geometry, e.g., large deformation, merging, and breaking. In the future, we will extend it to a full level set equation by additionally learning the level set function, so as to handle more complex interface problems and multi-medium flows in three dimensions.
    \item For shock--bubble interaction problems, the simulated physical time is still short and the accuracy has considerable room for improvement. In the future, we will develop a time-staged training strategy, which advances the simulation through successive temporal segments so as to enable long-time simulations, and incorporate the physical mechanism of shock--interface interaction to reduce the accumulation of errors over long simulation times.
    \item The scaling parameters $(\lambda, \mu)$ need to be tuned manually according to the ranges of the physical variables, which requires a priori knowledge of the problem. In the future, we will develop an adaptive selection strategy, so that the optimal scaling can be determined automatically for problems whose range of physical quantities is unknown in advance.
\end{itemize}

\section*{Acknowledgments}
The authors would like to thank Zhiqiang Zeng, Yibo Wang, Jianquan Wang, Guohang Pu for insightful discussions and constructive suggestions.

\appendix
\section*{Appendix A: Computational form of the governing equations}
\refstepcounter{section}
\label{app:pde}

In this appendix, we provide the explicit component-wise residual forms adopted in the numerical computation for the three flow configurations considered in this paper, namely the 1D plane flow ($n=1$, $\alpha=0$), the 1D cylindrically symmetric flow ($n=1$, $\alpha=1$), and the 2D plane flow ($n=2$, $\alpha=0$). In what follows, $U_{\sita}$ denotes the output of a sub-network, and $\rho$, $\boldsymbol u$, $p$ denote its components (the primitive variables), with the total energy $E = \frac{p+\Gamma B}{\Gamma-1} + \frac12\rho\,|\boldsymbol u|^2$. Subscripts denote partial derivatives, e.g., $\rho_t = \partial_t \rho$ and $\rho_{xx} = \partial_{xx} \rho$.

\begin{proposition}[Computational forms of the PDE residual]
\label{prop:app_pde}
For the three flow configurations, the PDE residual $\mathcal R^\nu(U,\Gamma,B)$ takes the following component-wise forms, where $\nu$ denotes the artificial viscosity coefficient:

\textbf{(1) 1D plane flow} ($n=1$, $\alpha=0$). In practice, we find that using the following computational form saves training time and achieves better accuracy:
\begin{equation}
\mathcal R^\nu(U,\Gamma,B)
=\begin{pmatrix}
\rho_t + u\,\rho_x + \rho\,u_x - \nu\,\rho_{xx}\\
\rho\,(u_t + u\,u_x) + p_x - \nu\,(\rho\,u_{xx} + 2\,\rho_x\,u_x)\\
p_t + u\,p_x + \Gamma(p+B)\,u_x - \nu\,\bigl(p_{xx} + \rho\,(\Gamma-1)\,u_x^2\bigr)
\end{pmatrix},
\end{equation}

\textbf{(2) 1D cylindrically symmetric flow} ($n=1$, $\alpha=1$). Directly using Eq.~\eqref{euler_pde} can easily lead to singularities at $x=0$. Therefore, we adopt the following computational form:
\begin{equation}
\mathcal R^\nu(U,\Gamma,B)
=\begin{pmatrix}
(x\rho)_t + (x\rho u)_x - \nu\,(x\rho)_{xx}\\[3pt]
(x\rho u)_t + \bigl(x(\rho u^2+p)\bigr)_x - p - \nu\,(x\rho u)_{xx}\\[3pt]
(xE)_t + \bigl(xu(E+p)\bigr)_x - \nu\,(xE)_{xx}
\end{pmatrix},
\end{equation}

\textbf{(3) 2D plane flow} ($n=2$, $\alpha=0$). Following a derivation similar to that of case (1), we obtain the following computational form:
\begin{equation}
\mathcal R^\nu(U,\Gamma,B)
=\begin{pmatrix}
\rho_t + u\,\rho_x + v\,\rho_y + \rho\,(u_x+v_y) - \nu\,(\rho_{xx}+\rho_{yy})\\[3pt]
\rho\,(u_t + u\,u_x + v\,u_y) + p_x - \nu\,\bigl(\rho\,(u_{xx}+u_{yy}) + 2(\rho_x u_x + \rho_y u_y)\bigr)\\[3pt]
\rho\,(v_t + u\,v_x + v\,v_y) + p_y - \nu\,\bigl(\rho\,(v_{xx}+v_{yy}) + 2(\rho_x v_x + \rho_y v_y)\bigr)\\[3pt]
p_t + u\,p_x + v\,p_y + \Gamma(p+B)\,(u_x+v_y) - \nu\,\bigl(p_{xx}+p_{yy} + \rho\,(\Gamma-1)\,|\nabla\boldsymbol u|^2\bigr)
\end{pmatrix},
\end{equation}
where $|\nabla\boldsymbol u|^2 = u_x^2 + u_y^2 + v_x^2 + v_y^2$.
\end{proposition}

\section*{Appendix B: Training costs of PINN-LPCDD}
\refstepcounter{section}
\label{app:runtime}

In this appendix, we summarize the training time, training speed, and GPU memory usage of PINN-LPCDD for each test case, as shown in Table~\ref{tab_runtime}, and discuss the computational cost for the 1D and 2D test cases, respectively.

\paragraph{1D cases.} The training cost of PINN-LPCDD is still not competitive with that of traditional high-precision methods: the training time is about 7 to 15 minutes, whereas computing the reference solution of test case (3B) with DG-$P^2$ \cite{wang2026conservation} on a uniform grid with $N_x=1000$ takes only 1.40 minutes.

\paragraph{2D cases.} The situation is different from 1D cases: the computational cost of PINN-LPCDD becomes comparable to that of traditional high-precision methods, and the numbers of spatial sampling points and time steps involved are also smaller than those of traditional methods. 

Due to GPU memory limitations, even when training on a coarse grid of $(N_t,N_x,N_y)=(16,128,64)$ in the space-time domain, we still have to resort to mini-batch training, and the final training time reaches nearly two hours for the test case (4B). 

For a fair comparison, the achieved accuracy should also be taken into account. The reference solution used in this paper is computed with DG-$P^1$ on a grid of $(N_x,N_y)=(400,400)$ with 16-core parallelization, which takes 101 minutes in total, only about 5 minutes less than the training time of our model on a single GPU. If the grid is coarsened to $(N_x,N_y)=(200,200)$ with 8-core parallelization, the computational time drops to 26 minutes, but the shock position is no longer accurately captured. In other words, at the accuracy level required for a reliable reference solution, traditional methods no longer hold a clear efficiency advantage over PINN-LPCDD. 

Note that this comparison is limited to the short-time simulation of the shock--bubble interaction problem considered in this paper; the efficiency for long-time simulations remains to be further investigated in future work.

\begin{table}[H]
\caption{Runtime, training speed, and GPU memory usage of PINN-LPCDD across all test cases. The training epoch is set to 5000 in Subsections~\ref{sec:air_air}--\ref{sec:air_water} and to 10000 for the shock-interface interaction problems in Subsection~\ref{sec:shock_interface}. The GPU memory refers to the cached/reserved memory of PyTorch.}\label{tab_runtime}
\small
\centering
\begin{tabular}{|c|ccc|cc|}
\hline
Case & (1A) & (1B) & (1C) & (2A) & (2B) \\
\hline
Runtime (min) & 7.47 & 7.59 & 7.54 & 7.54 & 7.40 \\
Speed (epoch/s) & 11.15 & 10.98 & 11.05 & 11.06 & 11.26 \\
GPU mem. (MB) & 2094 & 2090 & 2090 & 2090 & 2090 \\
\hline
Case & (3A) & (3B) & (3C) & (4A) & (4B) \\
\hline
Runtime (min) & 7.45 & 10.17 & 53.69 & 14.70 & 106.03 \\
Speed (epoch/s) & 11.18 & 8.19 & 1.55 & 11.34 & 1.57 \\
GPU mem. (MB) & 2094 & 2588 & 9156 & 2094 & 9158 \\
\hline
\end{tabular}
\end{table}

\bibliographystyle{unsrt}
\bibliography{references}

@article{clark2019three,
  title={Three-dimensional modeling and hydrodynamic scaling of {National Ignition Facility} implosions},
  author={Clark, Daniel S and Weber, Christopher R and Milovich, Jose L and Pak, Arthur E and Casey, Daniel T and Hammel, Bruce A and Ho, Darwin D and Jones, Ogden S and Koning, Joseph M and Kritcher, Andrea L and Marinak, Michael M and Mass{\'e}, Laurent P and Munro, David H and Patel, Mehul V and Patel, Pravesh K and Robey, Hayden F and Schroeder, C R and Sepke, Scott M and Edwards, M J},
  journal={Physics of Plasmas},
  volume={26},
  number={5},
  pages={050601},
  year={2019},
  publisher={AIP Publishing},
  doi={10.1063/1.5091449}
}

@article{klaseboer2005experimental,
  title={Experimental and numerical investigation of the dynamics of an underwater explosion bubble near a resilient/rigid structure},
  author={Klaseboer, E and Hung, KC and Wang, CW and Khoo, BC and Boyce, P and Debono, S and Charlier, Herve},
  journal={Journal of Fluid Mechanics},
  volume={537},
  pages={387--413},
  year={2005},
  publisher={Cambridge University Press}
}

@article{matula1997bjerknes,
  title={Bjerknes force and bubble levitation under single-bubble sonoluminescence conditions},
  author={Matula, Thomas J and Cordry, Sean M and Roy, Ronald A and Crum, Lawrence A},
  journal={The Journal of the Acoustical Society of America},
  volume={102},
  number={3},
  pages={1522--1527},
  year={1997},
  publisher={Acoustical Society of America}
}

@article{raissi2019physics,
  title={{Physics-informed neural networks: A deep learning framework for solving forward and inverse problems involving nonlinear partial differential equations}},
  author={Raissi, Maziar and Perdikaris, Paris and Karniadakis, George E},
  journal={Journal of Computational physics},
  volume={378},
  pages={686--707},
  year={2019},
  publisher={Elsevier}
}

@article{feng2020characteristic,
  title={{A characteristic-featured shock wave indicator for conservation laws based on training an artificial neuron}},
  author={Feng, Yiwei and Liu, Tiegang and Wang, Kun},
  journal={Journal of Scientific Computing},
  volume={83},
  pages={1--34},
  year={2020},
  publisher={Springer}
}

@article{cockburn1989tvb,
  title={{TVB Runge-Kutta local projection discontinuous Galerkin finite element method for conservation laws. II. General framework}},
  author={Cockburn, Bernardo and Shu, Chi-Wang},
  journal={Mathematics of computation},
  volume={52},
  number={186},
  pages={411--435},
  year={1989}
}

@article{lu2021deepxde,
  title={{DeepXDE: A deep learning library for solving differential equations}},
  author={Lu, Lu and Meng, Xuhui and Mao, Zhiping and Karniadakis, George Em},
  journal={SIAM review},
  volume={63},
  number={1},
  pages={208--228},
  year={2021},
  publisher={SIAM}
}

@article{liu2024discontinuity,
  title={{Discontinuity computing using physics-informed neural networks}},
  author={Liu, Li and Liu, Shengping and Xie, Hui and Xiong, Fansheng and Yu, Tengchao and Xiao, Mengjuan and Liu, Lufeng and Yong, Heng},
  journal={Journal of Scientific Computing},
  volume={98},
  number={1},
  pages={22},
  year={2024},
  publisher={Springer}
}

@article{ferrer2024gradient,
  title={{Gradient-annihilated PINNs for solving Riemann problems: Application to relativistic hydrodynamics}},
  author={Ferrer-Sanchez, Antonio and Martin-Guerrero, Jose D and de Austri-Bazan, Roberto Ruiz and Torres-Forne, Alejandro and Font, Jose A},
  journal={Computer Methods in Applied Mechanics and Engineering},
  volume={424},
  pages={116906},
  year={2024},
  publisher={Elsevier}
}

@article{ray2018artificial,
  title={{An artificial neural network as a troubled-cell indicator}},
  author={Ray, Deep and Hesthaven, Jan S},
  journal={Journal of computational physics},
  volume={367},
  pages={166--191},
  year={2018},
  publisher={Elsevier}
}

@article{feng2021characteristic,
  title={{A characteristic-featured shock wave indicator on unstructured grids based on training an artificial neuron}},
  author={Feng, Yiwei and Liu, Tiegang},
  journal={Journal of Computational Physics},
  volume={443},
  pages={110446},
  year={2021},
  publisher={Elsevier}
}

@article{jiang1996efficient,
  title={{Efficient implementation of weighted ENO schemes}},
  author={Jiang, Guang-Shan and Shu, Chi-Wang},
  journal={Journal of computational physics},
  volume={126},
  number={1},
  pages={202--228},
  year={1996},
  publisher={Elsevier}
}

@techreport{reed1973triangular,
  title={{Triangular mesh methods for the neutron transport equation}},
  author={Reed, William H and Hill, Thomas R},
  year={1973},
  institution={Los Alamos Scientific Lab., N. Mex.(USA)}
}

@article{cockburn1991runge,
  title={{The Runge-Kutta local projection-discontinuous-Galerkin finite element method for scalar conservation laws}},
  author={Cockburn, Bernardo and Shu, Chi-Wang},
  journal={ESAIM: Mathematical Modelling and Numerical Analysis},
  volume={25},
  number={3},
  pages={337--361},
  year={1991},
  publisher={EDP Sciences}
}

@article{he2023hodg,
  title={{HODG: high-order discontinuous Galerkin methods for solving compressible Euler and Navier-Stokes equations-an open-source component-based development framework}},
  author={He, Xiaofeng and Wang, Kun and Liu, Tiegang and Feng, Yiwei and Zhang, Bin and Yuan, Weixiong and Wang, Xiaojun},
  journal={Computer Physics Communications},
  volume={286},
  pages={108660},
  year={2023},
  publisher={Elsevier}
}

@article{abbasi2025challenges,
  title={Challenges and advancements in modeling shock fronts with physics-informed neural networks: A review and benchmarking study},
  author={Abbasi, Jassem and Jagtap, Ameya D and Moseley, Ben and Hiorth, Aksel and Andersen, P{\aa}l {\O}steb{\o}},
  journal={arXiv preprint arXiv:2503.17379},
  year={2025}
}

@article{friedrich1998weighted,
  title={Weighted essentially non-oscillatory schemes for the interpolation of mean values on unstructured grids},
  author={Friedrich, Oliver},
  journal={Journal of computational physics},
  volume={144},
  number={1},
  pages={194--212},
  year={1998},
  publisher={Elsevier}
}

@article{hu1999weighted,
  title={Weighted essentially non-oscillatory schemes on triangular meshes},
  author={Hu, Changqing and Shu, Chi-Wang},
  journal={Journal of Computational Physics},
  volume={150},
  number={1},
  pages={97--127},
  year={1999},
  publisher={Elsevier}
}

@article{nguyen2022machine,
  title={Machine learning changes the rules for flux limiters},
  author={Nguyen-Fotiadis, Nga and McKerns, Michael and Sornborger, Andrew},
  journal={Physics of Fluids},
  volume={34},
  number={8},
  pages={085136},
  year={2022},
  publisher={AIP Publishing}
}

@article{morand2024deep,
  title={Deep learning of first-order nonlinear hyperbolic conservation law solvers},
  author={Morand, Victor and M{\"u}ller, Nils and Weightman, Ryan and Piccoli, Benedetto and Keimer, Alexander and Bayen, Alexandre M},
  journal={Journal of Computational Physics},
  volume={511},
  pages={113114},
  year={2024},
  publisher={Elsevier}
}

@article{bois2024optimal,
  title={An optimal control deep learning method to design artificial viscosities for discontinuous {Galerkin} schemes},
  author={Bois, L{\'e}o and Franck, Emmanuel and Navoret, Laurent and Vigon, Vincent},
  journal={Journal of Scientific Computing},
  volume={101},
  number={3},
  pages={70},
  year={2024},
  publisher={Springer}
}

@article{yu2022data,
  title={A data-driven shock capturing approach for discontinuous {Galerkin} methods},
  author={Yu, Jian and Hesthaven, Jan S},
  journal={Computers \& Fluids},
  volume={245},
  pages={105592},
  year={2022},
  publisher={Elsevier}
}

@article{jagtap2020conservative,
  title={Conservative physics-informed neural networks on discrete domains for conservation laws: Applications to forward and inverse problems},
  author={Jagtap, Ameya D and Kharazmi, Ehsan and Karniadakis, George Em},
  journal={Computer Methods in Applied Mechanics and Engineering},
  volume={365},
  pages={113028},
  year={2020},
  publisher={Elsevier}
}

@article{jagtap2020extended,
  title={Extended physics-informed neural networks ({XPINNs}): A generalized space-time domain decomposition based deep learning framework for nonlinear partial differential equations},
  author={Jagtap, Ameya D and Karniadakis, George Em},
  journal={Communications in Computational Physics},
  volume={28},
  number={5},
  pages={2002--2041},
  year={2020},
  publisher={Brown Univ., Providence, RI (United States)}
}

@article{mao2023physics,
  title={Physics-informed neural networks with residual/gradient-based adaptive sampling methods for solving partial differential equations with sharp solutions},
  author={Mao, Zhiping and Meng, Xuhui},
  journal={Applied Mathematics and Mechanics},
  volume={44},
  number={7},
  pages={1069--1084},
  year={2023},
  publisher={Springer}
}

@article{urban2025riemann,
  title={An approximate {Riemann} solver approach in physics-informed neural networks for hyperbolic conservation laws},
  author={Urb{\'a}n, Jorge F and Pons, J A},
  journal={Physics of Fluids},
  volume={37},
  number={9},
  pages={096114},
  year={2025},
  publisher={AIP Publishing}
}

@article{jnini2026curvature,
  title={Curvature-Aware Optimization for High-Accuracy Physics-Informed Neural Networks},
  author={Jnini, Anas and Kiyani, Elham and Shukla, Khemraj and Urb{\'a}n, Jorge F and Daryakenari, Nazanin Ahmadi and Muller, Johannes and Zeinhofer, Marius and Karniadakis, George Em},
  journal={arXiv preprint arXiv:2604.05230},
  year={2026}
}

@article{lei2025discontinuity,
  title={Discontinuity-aware {KAN}-based physics-informed neural networks},
  author={Lei, Guoqiang and Exposito, D and Mao, Xuerui},
  journal={arXiv preprint arXiv:2507.08338},
  year={2025}
}

@article{zeng2026scinn,
  title={{SCINN: Solution Character-Informed Neural Network for Scalar Conservation Laws}},
  author={Zeng, Weiheng and Lu, Ruoxi and Wang, Kun and Liu, Tiegang},
  journal={Journal of Machine Learning},
  volume={5},
  number={3},
  pages={525--559},
  year={2026}
}

@article{xu2025preprocessing,
  title={On the preprocessing of physics-informed neural networks: How to better utilize data in fluid mechanics},
  author={Xu, Shengfeng and Dai, Yuanjun and Yan, Chang and Sun, Zhenxu and Huang, Renfang and Guo, Dilong and Yang, Guowei},
  journal={Journal of Computational Physics},
  volume={528},
  pages={113837},
  year={2025},
  publisher={Elsevier}
}

@article{peyvan2024riemann,
  title={{RiemannONets}: Interpretable neural operators for {Riemann} problems},
  author={Peyvan, Ahmad and Oommen, Vivek and Jagtap, Ameya D and Karniadakis, George Em},
  journal={Computer Methods in Applied Mechanics and Engineering},
  volume={426},
  pages={116996},
  year={2024},
  publisher={Elsevier}
}

@article{liu2026data,
  title={Data-Free physics-informed neural networks for modeling compressible multiphase flows},
  author={Liu, Rui and Zhao, Zitong and Rong, Jili},
  journal={International Journal of Multiphase Flow},
  volume={196},
  pages={105589},
  year={2026},
  publisher={Elsevier}
}

@article{buhendwa2021inferring,
  title={Inferring incompressible two-phase flow fields from the interface motion using physics-informed neural networks},
  author={Buhendwa, A B and Adami, S and Adams, N A},
  journal={Machine Learning with Applications},
  volume={4},
  pages={100029},
  year={2021},
  publisher={Elsevier}
}

@article{zhu2023physics,
  title={Physics-informed neural networks for solving dynamic two-phase interface problems},
  author={Zhu, Xingwen and Hu, Xiaozhe and Sun, Pengtao},
  journal={SIAM Journal on Scientific Computing},
  volume={45},
  number={6},
  pages={A2912--A2944},
  year={2023},
  publisher={SIAM}
}

@article{radhakrishnan2024distributed,
  title={Distributed physics-informed machine learning strategies for two-phase flows},
  author={Radhakrishnan, G and Pattamatta, A and Srinivasan, B},
  journal={International Journal of Multiphase Flow},
  volume={177},
  pages={104861},
  year={2024},
  publisher={Elsevier}
}

@article{lax1954weak,
  title={Weak solutions of nonlinear hyperbolic equations and their numerical computation},
  author={Lax, Peter D},
  journal={Communications on pure and applied mathematics},
  volume={7},
  number={1},
  pages={159--193},
  year={1954},
  publisher={Wiley Online Library}
}

@article{sod1978survey,
  title={A survey of several finite difference methods for systems of nonlinear hyperbolic conservation laws},
  author={Sod, Gary A},
  journal={Journal of computational physics},
  volume={27},
  number={1},
  pages={1--31},
  year={1978},
  publisher={Elsevier}
}

@article{allaire2002five,
  title={A five-equation model for the simulation of interfaces between compressible fluids},
  author={Allaire, Gr{\'e}goire and Clerc, S{\'e}bastien and Kokh, Samuel},
  journal={Journal of Computational Physics},
  volume={181},
  number={2},
  pages={577--616},
  year={2002},
  publisher={Elsevier}
}

@article{osher1988fronts,
  title={Fronts propagating with curvature-dependent speed: algorithms based on {Hamilton-Jacobi} formulations},
  author={Osher, Stanley and Sethian, James A},
  journal={Journal of Computational Physics},
  volume={79},
  number={1},
  pages={12--49},
  year={1988},
  publisher={Elsevier}
}

@book{sethian1999level,
  title={Level set methods and fast marching methods: evolving interfaces in computational geometry, fluid mechanics, computer vision, and materials science},
  author={Sethian, James A},
  year={1999},
  publisher={Cambridge University Press}
}

@article{fedkiw1999non,
  title={A non-oscillatory {Eulerian} approach to interfaces in {Multimaterial} flows (the ghost fluid method)},
  author={Fedkiw, Ronald P and Aslam, Tariq and Merriman, Barry and Osher, Stanley},
  journal={Journal of Computational Physics},
  volume={152},
  number={2},
  pages={457--492},
  year={1999},
  publisher={Elsevier}
}

@article{shyue2006efficient,
  title={An efficient shock-capturing algorithm for compressible multicomponent problems},
  author={Shyue, Keh-Ming},
  journal={Journal of Computational Physics},
  volume={142},
  number={1},
  pages={208--242},
  year={1998},
  publisher={Elsevier}
}

@article{xu2024physics,
  title={Physics-constrained neural networks as multi-material {Riemann} solvers for compressible two-gas simulations},
  author={Xu, Liang and Liu, Ziyan and Feng, Yiwei and Liu, Tiegang},
  journal={Journal of Computational Science},
  volume={78},
  pages={102261},
  year={2024},
  publisher={Elsevier}
}

@article{xu2025unsupervised,
  title={Unsupervised neural-network solvers for multi-material {Riemann} problems},
  author={Xu, Liang and Liu, Ziyan and Feng, Yiwei and Liu, Tiegang},
  journal={Computer Physics Communications},
  volume={308},
  pages={109470},
  year={2025},
  publisher={Elsevier}
}

@article{zhou2024self,
  title={Self-adaptive and time divide-and-conquer physics-informed neural networks for two-phase flow simulations using interface tracking methods},
  author={Zhou, Wen and Miwa, Shuichiro and Okamoto, Koji},
  journal={Physics of Fluids},
  volume={36},
  number={7},
  pages={073305},
  year={2024},
  publisher={AIP Publishing}
}

@article{song2026physics,
  title={Physics-informed neural networks for shock capturing in inviscid flows around an airfoil},
  author={Song, Jiahao and Cao, Wenbo and Zhang, Weiwei},
  journal={arXiv preprint arXiv:2607.08130},
  year={2026}
}

@article{liu2003ghost,
  title={Ghost fluid method for strong shock impacting on material interface},
  author={Liu, T G and Khoo, B C and Yeo, KS},
  journal={Journal of computational physics},
  volume={190},
  number={2},
  pages={651--681},
  year={2003},
  publisher={Elsevier}
}

@article{liu2005ghost,
  title={The ghost fluid method for compressible gas--water simulation},
  author={Liu, Tiegang and Khoo, Boo Cheong and Wang, C W},
  journal={Journal of Computational Physics},
  volume={204},
  number={1},
  pages={193--221},
  year={2005},
  publisher={Elsevier}
}

@inproceedings{rahaman2019spectral,
  title={On the spectral bias of neural networks},
  author={Rahaman, Nasim and Baratin, Aristide and Arpit, Devansh and Draxler, Felix and Lin, Min and Hamprecht, Fred and Bengio, Yoshua and Courville, Aaron},
  booktitle={International Conference on Machine Learning},
  pages={5301--5310},
  year={2019},
  organization={PMLR}
}

@inproceedings{tancik2020fourier,
  title={Fourier features let networks learn high frequency functions in low dimensional domains},
  author={Tancik, Matthew and Srinivasan, Pratul P and Mildenhall, Ben and Fridovich-Keil, Sara and Raghavan, Nithin and Singhal, Utkarsh and Ramamoorthi, Ravi and Barron, Jonathan T and Ng, Ren},
  booktitle={Advances in Neural Information Processing Systems},
  volume={33},
  pages={7537--7547},
  year={2020}
}

@article{haas1987interaction,
  title={Interaction of weak shock waves with cylindrical and spherical gas inhomogeneities},
  author={Haas, J-F and Sturtevant, Bradford},
  journal={Journal of Fluid Mechanics},
  volume={181},
  pages={41--76},
  year={1987},
  publisher={Cambridge University Press}
}

@book{toro2013riemann,
  title={Riemann solvers and numerical methods for fluid dynamics: a practical introduction},
  author={Toro, Eleuterio F},
  year={2013},
  publisher={Springer Science \& Business Media}
}

@inproceedings{vyas2025soap,
  title={{SOAP}: Improving and stabilizing shampoo using {Adam} for language modeling},
  author={Vyas, Nikhil and Morwani, Depen and Zhao, Rosie and Shapira, Itai and Brandfonbrener, David and Janson, Lucas and Kakade, Sham},
  booktitle={International Conference on Learning Representations},
  year={2025}
}

@article{zhang2025physics,
  title={Physics-informed neural networks for multi-material high-accuracy {Riemann} solver in {Eulerian} framework},
  author={Zhang, Sen and Zhong, Yanxu and Yang, Xi and Guo, Xiao-Wei and Zhang, Zhuo and Wang, Wei and Yang, Canqun},
  journal={Physics of Fluids},
  volume={37},
  number={8},
  pages={082110},
  year={2025},
  publisher={AIP Publishing}
}

@article{zhai2026physics,
  title={Physics-informed neural networks for solving two-phase flow problems with moving interfaces},
  author={Zhai, Qijia and Sun, Pengtao and Xie, Xiaoping and Zhu, Xingwen and Zhang, Chen-Song},
  journal={arXiv preprint arXiv:2604.00948},
  year={2026}
}

@article{glimm1998three,
  title={Three-dimensional front tracking},
  author={Glimm, James and Grove, John W and Li, Xiaolin and Shyue, Keh-Ming and Zeng, Yanan and Zhang, Qiang},
  journal={SIAM Journal on Scientific Computing},
  volume={19},
  number={3},
  pages={703--727},
  year={1998},
  publisher={SIAM}
}

@article{tryggvason2001front,
  title={A front-tracking method for the computations of multiphase flow},
  author={Tryggvason, Gr{\'e}tar and Bunner, Bernard and Esmaeeli, Alain and Juric, Damir and Al-Rawahi, Nasser and Tauber, William and Han, Jang and Nas, Sam and Jan, Yuriko-Jane},
  journal={Journal of Computational Physics},
  volume={169},
  number={2},
  pages={708--759},
  year={2001},
  publisher={Elsevier}
}

@article{wang2026conservation,
  title={{Conservation Error Correction for MGFM-Type Methods Within the DG Framework for Axisymmetric Compressible Multimedium Flows}},
  author={Wang, Hao and Zhang, Xiaotao and Feng, Chengliang and Liu, Tiegang and Zeng, Zhiqiang},
  journal={Communications in Computational Physics},
  year={2026},
  doi={10.4208/cicp.OA-2025-0305}
}

@article{wassing2025physics,
  title={Physics-informed neural networks for inviscid transonic flows around an airfoil},
  author={Wassing, Simon and Langer, Stefan and Bekemeyer, Philipp},
  journal={Physics of Fluids},
  volume={37},
  number={8},
  pages={086169},
  year={2025},
  publisher={AIP Publishing}
}

@article{ding2026adaptive,
  title={Adaptive weighted multi-physics-informed neural network for high-accuracy shock wave capturing},
  author={Ding, Shuaibing and Lei, Juanmian and Xu, Liang and Zhang, Boqian and Sun, Guoyou and Guo, Jian},
  journal={International Journal of Computational Methods},
  volume={23},
  number={03},
  pages={2550063},
  year={2026},
  publisher={World Scientific}
}

@article{xiao2022hybrid,
  title={{A Hybrid Multilayer Perceptron-Radial Basis Function (HMLP-RBF) Neural Network for Solving Hyperbolic Conservation Laws}},
  author={Xiao, Yang and Yang, Liming and Yuan, Haizhuan and Shu, Chang},
  journal={SN Computer Science},
  volume={3},
  number={6},
  pages={490},
  year={2022},
  publisher={Springer},
  doi={10.1007/s42979-022-01413-5}
}

@article{li2026adaptive,
  title={{Adaptive Regularized Hybrid Physics-Informed Neural Networks for Large-Gradient Problems}},
  author={Li, Tingjie and Zheng, Supei and Hu, Fengli and Lu, Xiaoli and Xu, Doudou and Cheng, Xiaohan and Song, Xueli},
  journal={Physics of Fluids},
  volume={38},
  number={8},
  pages={086106},
  year={2026},
  publisher={AIP Publishing},
  doi={10.1063/5.0338767}
}

@inproceedings{glorot2010understanding,
  title={Understanding the difficulty of training deep feedforward neural networks},
  author={Glorot, Xavier and Bengio, Yoshua},
  booktitle={International Conference on Artificial Intelligence and Statistics},
  pages={249--256},
  year={2010}
}

\end{document}